\documentclass[10pt,journal,compsoc]{IEEEtran}

\usepackage{xspace}
\usepackage{color}

\ifCLASSOPTIONcompsoc
  \usepackage[nocompress]{cite}
\else
  \usepackage{cite}
\fi

\ifCLASSINFOpdf
  \usepackage[pdftex]{graphicx}
  \graphicspath{{Pictures/} {AuthorPics/} {ImgTemplate/}}
  \DeclareGraphicsExtensions{.pdf,.jpeg,.png}
\else

\fi

\usepackage{amsmath,amssymb,amsfonts}

\usepackage{algorithmic}
\usepackage{textcomp}

\usepackage{url}

\usepackage[inline]{enumitem}

\usepackage{tikz}
\usetikzlibrary{calc,arrows,shapes,positioning,patterns,fit,automata,decorations.markings}

\newcommand{\ie}{i.e.,\xspace}
\newcommand{\eg}{e.g.,\xspace}

\newcommand{\FPGA}{\ensuremath{\mathrm{FPGA}}\xspace}

\newcommand{\SW}{\ensuremath{\mathrm{SW}}\xspace}

\newcommand{\BS}{\ensuremath{\mathrm{BS}}\xspace}

\newenvironment{note1}[1]{\small
   \begin{center}
   \begin{tabular}{|p{7cm}|}\hline
   \bf{#1}: \sf
}{
  \\\hline
   \end{tabular}
   \end{center}
}

\newcommand{\STORE}{\ensuremath{\mathrm{STORE}}\xspace}
\newcommand{\STOREext}{Software Application Store\xspace}
\newcommand{\PGW}{\ensuremath{\mathrm{PG}}\xspace}
\newcommand{\PGWext}{Payment Gateway\xspace}
\newcommand{\SWP}{\ensuremath{\mathrm{SWP}}\xspace}
\newcommand{\SWPext}{Software Provider\xspace}
\newcommand{\HWV}{\ensuremath{\mathrm{HWV}}\xspace}
\newcommand{\HWVext}{Hardware Vendor\xspace}

\newcommand{\EU}{\ensuremath{\mathrm{EU}}\xspace}
\newcommand{\EUext}{End User\xspace}
\newcommand{\EUDext}{End User Device\xspace}

\newcommand{\HAP}{\ensuremath{\mathrm{HAP}}\xspace}
\newcommand{\HAPext}{Hardware Acceleration Platform\xspace}
\newcommand{\IDHAP}{\ensuremath{\mathrm{id_{H\!A\!P}}}\xspace}
\newcommand{\KHAP}{\ensuremath{\mathrm{K_{H\!A\!P}}}\xspace}
\newcommand{\HAPkey}{\HAP key\xspace}

\newcommand{\BSenc}{\ensuremath{\{\mathrm{BS}\}_{\KHAP}}\xspace}

\ifCLASSOPTIONcompsoc
  \usepackage[nocompress]{cite}
\else
  \usepackage{cite}
\fi

\ifCLASSINFOpdf
  \usepackage[pdftex]{graphicx}
\else
  \usepackage[dvips]{graphicx}
\fi

\usepackage{amsmath}

\usepackage{algorithmic}
\usepackage{algorithm}

\ifCLASSOPTIONcompsoc
  \usepackage[caption=false,font=footnotesize,labelfont=sf,textfont=sf]{subfig}
\else
  \usepackage[caption=false,font=footnotesize]{subfig}
\fi

\usepackage{url}

\begin{document}
%
\title{Securing Soft IP Cores in FPGA based Reconfigurable Mobile Heterogeneous Systems}

\author{Alberto~Carelli,~\IEEEmembership{Student Member,~IEEE,}
		Cataldo~Basile, ~\IEEEmembership{Member,~IEEE,}
		Alessandro~Savino, ~\IEEEmembership{Member,~IEEE,}
		Alessandro~Vallero, ~\IEEEmembership{Member,~IEEE,}
        and~Stefano~Di~Carlo~\IEEEmembership{Senior Member,~IEEE}
\IEEEcompsocitemizethanks{\IEEEcompsocthanksitem A. Carelli, C. Basile, A. Savino, A. Vallero and S. Di Carlo are with the Department
of Control and Computer Engineering, Politecnico di Torino, Torino,
Italy, 10129.\protect\\
E-mail: {name.surname}@polito.it
}
}

\IEEEtitleabstractindextext{%
\begin{abstract}
The mobile application market is rapidly growing and changing, offering always brand new software to install in increasingly powerful devices. 
Mobile devices become pervasive and more heterogeneous, embedding latest technologies such as multicore architectures, special-purpose circuits and reconfigurable logic.
In a future mobile market scenario reconfigurable systems are employed to provide high-speed functionalities to assist execution of applications.
However, new security concerns are introduced. In particular, protecting the Intellectual Property of the exchanged soft IP cores is a serious concern.
The available techniques for preserving integrity, confidentiality and authenticity suffer from the limitation of heavily relying onto the system designer.
In this paper we propose two different protocols suitable for the secure deployment of soft IP cores in FPGA-based mobile heterogeneous systems where multiple independent actors are involved: a simple scenario requiring trust relationship between entities, and a more complex scenario where no trust relationship exists through adoption of the Direct Anonymous Attestation protocol. 
Finally, we provide a prototype implementation of the proposed architectures.

\end{abstract}

\begin{IEEEkeywords}
FPGA, Bitstream confidentiality, Bitstream deployment, Bitstream security, Direct Anonymous Attestation, Heterogeneous systems, Hardware security
\end{IEEEkeywords}}

\maketitle

\IEEEdisplaynontitleabstractindextext
\IEEEpeerreviewmaketitle

\section{Introduction}
\label{sec:introduction}
The current mobile application market is constantly changing due to the presence of new devices and platforms which emerge quite frequently \cite{Albright:fk}. 
The changes affect the business environment creating a demand but also an opportunity for the rapid introduction of new technological solutions. 
Players in the growing business landscape cannot ignore this opportunity, since mobile technology will eventually have a role in most digital products and services.

Several well-known companies participate in this scenario: the first to define the market was Apple which launched, in 2007, the iPhone and later, in 2008, its distribution platform App Store.
Subsequently, other players and device manufacturers joined the mobile application market with their devices, operating systems and software distribution platforms.
Currently, the distribution of mobile applications spans over 300 application stores worldwide, including device manufacturers, platform providers, mobile operators. Well-known examples are Google Play Store (previously known as Android Market), Apple App Store, Windows Phone Store, Opera Mobile Store, etc. \cite{appstorelist}.
Sales, finalized through a payment gateway, and downloads of mobile apps are skyrocketing too.
According to a recent survey by iResearch, in 2018, global mobile app revenues amounted to over 365 billion U.S. dollars. In 2023, mobile apps are projected to generate more than 935 billion U.S. dollars in revenues via paid downloads from app stores and in-app advertising \cite{iresearch}.

Several motivations are at the base of the continuous growth of this phenomenon. 
Certainly, the hardware improvement is the key factor that keeps pushing applications toward new levels of pervasiveness, which allows an ever increasing computational power of mobile platforms.
Second, heterogeneous computing has been the leading technology that allowed moving towards new generations of mobile devices. For example, combining multi-core processors with Graphics Processing Units (GPUs) and other types of hardware accelerators is having a huge impact on device performance.
Therefore, System-on-Chip developers are increasingly looking at alternative architectural solutions to increase the computational power, while optimizing additional parameters such as power consumption. 

In this landscape, reconfigurable computing is a promising solution.
As shown in \cite{Ke-fei:2010ly, Surratt:2005ve}, programmable system platforms embedding Field-Programmable Gate-Arrays (FPGAs) might be a valuable solution where frequent and remote upgrades are necessary, thus also for embedded applications.
Moreover, some FPGA-based systems provide Dynamic Partial Re-configuration (DPR), which allows the run-time update of selected portions of an FPGA without disrupting the rest of the system.
DPR allows the creation of new application scenarios \cite{Ahmad2009a, Dunham2009, SDCDPR6645549} and it has already been used in the mobile application market \cite{7169365, 7406067, Perera:2015:AFR:2889287}. 
%
Hardware vendors are responsible for manufacturing the FPGA devices and sell them to their customers or retailers.
In FPGAs it is possible to (re)configure logic resources  by controlling the interconnection among different logic gates. 
The hardware logic blocks implementing specific functions compose an Intellectual Property (IP) soft IP core. 
A soft IP core is an independent and reusable module that can be instantiated in the reconfigurable fabric.
One or more soft IP cores are described by a \emph{bitstream} file, which configures the FPGA (or just a portion, if DPR is supported). The project of a soft IP core is defined by the system designer.
The exploitation of reconfigurable hardware enables a new mobile application scenario, where a reconfigurable device can be programmed at run-time to assist the execution of a software application by means of application-specific computational cores. The applications can employ hardware capabilities on-demand using ad-hoc computational resources optimized for the various system's aspects (\eg power consumption of the whole system).
Mobile applications like games, audio/video processing, secure communications are good candidates to benefit from providing application-specific hardware acceleration cores deployed together with the software application.

However, this new application paradigm opens up concerns in the security domain. 
As an example, an adversary that is able to intercept a bitstream of an hardware Intellectual Property can try to extract sensitive information or steal the property, thus leading to IP infringements.
Also disclosure to the public domain or unauthorized sales to earn unfair profit are possible, thus violating the confidentiality of the hardware block. 
Considering the hardware, 
an attacker may also produce intentional alterations to the hardware block by injecting malicious code that may either corrupt the correct behavior of the software application or compromise the whole end-users system, e.g., by introducing security flaws which could later be exploited.

Therefore, the hardware description of soft IP cores should never be sent across unprotected network links.
FPGA manufacturers have already provided techniques, such as bitstream encryption and authentication, which try to address these issues. 
Indeed, they fit a simple scenario where the application developer and designer is the only entity entrusted to produce reconfigurable hardware descriptions to be delivered to a remote system. 
Certainly, they don't fit the current very dynamic and heterogeneous scenario of the mobile application market.
Moreover, these methods may require an excessive involvement and trust in the manufacturers, which usually embed secrets in their hardware that can eventually be exploited. Even worse, it requires to trust the whole manufacturers' supply chain as well, which has become an issue with the delocalization of the production. 

In this work, we tackle a much more challenging and realistic situation that involves several independent parties participating in the development and distribution of applications to be executed on heterogeneous mobile platforms embedding an FPGA as reconfigurable hardware. These parties include: the end users, the application stores, the software providers and the reconfigurable hardware vendors. 
In this scenario, a single reconfigurable hardware resource can be shared by several applications from different vendors with guarantees on the integrity and confidentiality of the provided hardware cores. 

This paper extends a preliminary published work by adding tightening security requirements and developing a more complex scenario that avoids trust relationships among the involved entities \cite{aqtr2018}.

This paper addresses the security threats introduced by two types of adversaries: 
\begin{enumerate*}[label=(\roman*)]
\item remote adversaries acting on the communication channels between the application providers and the devices (Man-in-the-Middle), and 
\item local adversaries with physical access to the system (Man-at-the-End). 
\end{enumerate*}

We describe the security services for the envisioned infrastructure by taking into account three aspects: 
\begin{enumerate*}[label=(\roman*)]
\item the hardware resources and the system architecture to implement the required security primitives, 
\item the high-level software infrastructure needed to implement the required communication protocols, and
\item the high-level entrusting policies required among the involved entities.
\end{enumerate*}

Instead of resorting to ad-hoc technology to tackle the adversaries and the security issues, we exploit the idea of another mainstream initiative, \ie the Direct Anonymous Attestation (DAA) protocol \cite{BrickellCC04}.
Currently, the DAA protocol has been standardized by the Trusted Computing Group (TCG) \cite{TPMSPECS} and it is supported in ad-hoc chips like the Trusted Platform Module (TPM).
Even if we do not necessarily propose the use of TCG-specified TPM chips, we follow the progresses in this field employing reconfigurable hardware to achieve the same features.
This also means that there will be more people researching for flaws, as the impact of attacks becomes greater thus also the impact of publication of such attacks is more likely to reach a higher visibility.

The remainder of this paper is organized as follows: 
Section \ref{sec:assumptions} defines the assumptions and the security requirements to be satisfied, as well as the roles of the actors playing in the two scenarios here considered.
Section \ref{sec:related-works} provides an overview of the previous works carried out in the perspective of the security requirements.
In Section \ref{sec:protocols}, we detail the protocol to securely transfer the bitstream.
We discuss, in Section \ref{sec:hwarch}, the internal architecture of the employed device, providing details for the software implementation of the considered players.
In Section \ref{sec:security_analysis},  we 
analyze the security of the protocols considering their security requirements against possible attacks.
Finally, Section \ref{sec:conclusion}, concludes the paper.

\section{Assumptions, models and requirements}
\label{sec:assumptions}

This section presents our model by describing the involved actors and the security and functional requirements to be satisfied.
Moreover, it characterizes the two scenarios we assume in this paper, which depict very common situations for the deployment of mobile applications.

\begin{figure}[!htb]
	\centering
	\includegraphics[width=0.8\columnwidth]{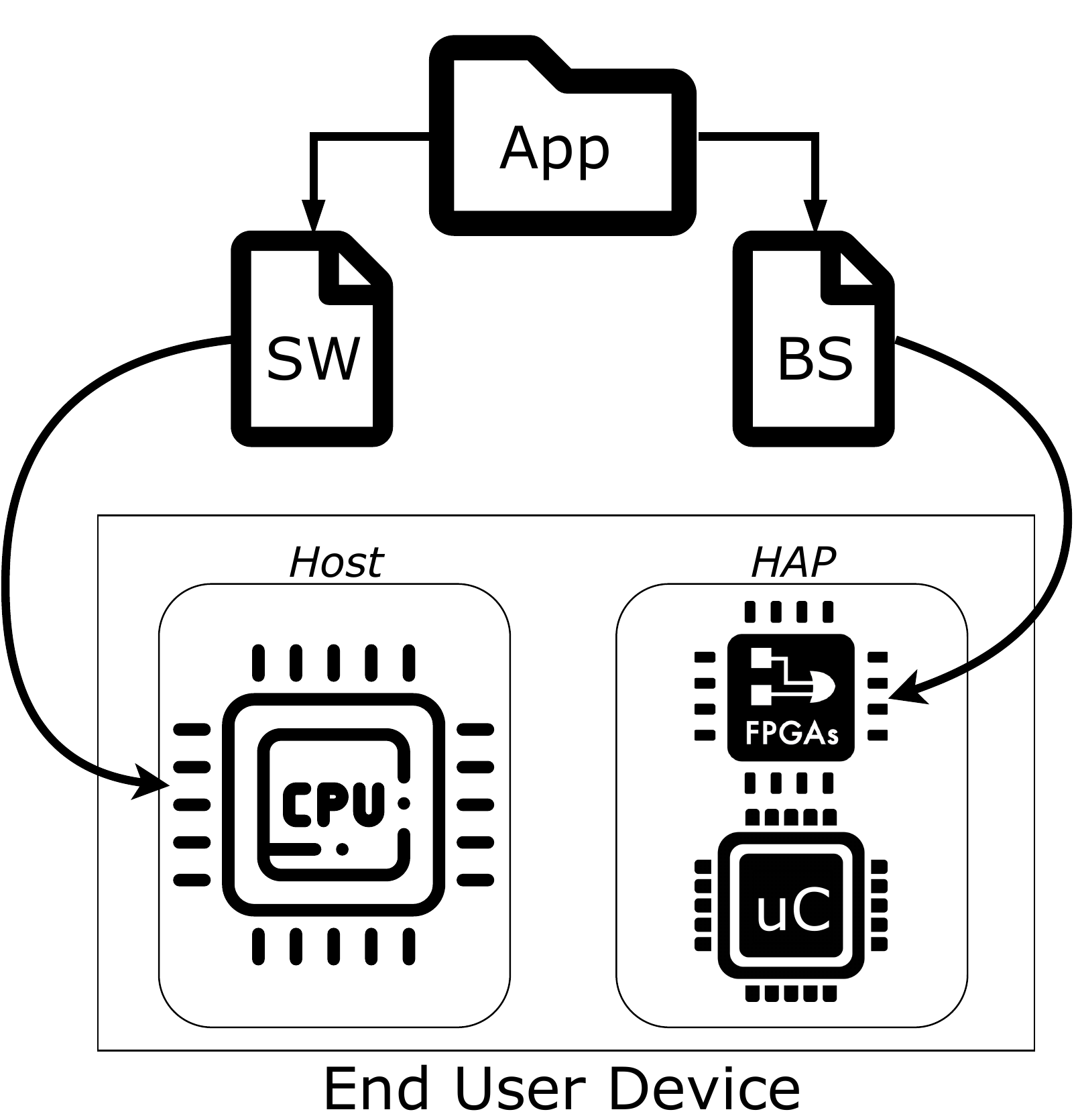}
	\caption{Mobile application paradigm considered mapped onto the (simplified) architecture of the End User Device}
	\label{fig:appmodel}
\end{figure}

The \textit{application} is the object to be securely exchanged among all involved parties. 
As shown in Fig.~\ref{fig:appmodel}, in this paper we consider applications composed of two portions: 
\begin{enumerate*}[label=(\roman*)]
\item the executable code (\SW), and 
\item an \emph{FPGA Bitstream file} (\BS).
\end{enumerate*}

The application is executed on the \EUDext, which embeds a general purpose CPU executing the \SW and a FPGA-based \HAPext (\HAP).
\HAP contains the reconfigurable logic, \ie the FPGA, and a microcontroller used to securely load and store a bitstream. 

The bitstream describes one or more soft IP cores that complement the application (e.g., required to improve its performance). These soft IP cores are dynamically configured in the FPGA every time the application is executed.


\subsection{Actors}
\label{sec:assumptions:entities}
\label{sec:assumptions:actors}

\begin{figure*}[!htb]
	\centering
	\includegraphics[width=2\columnwidth]{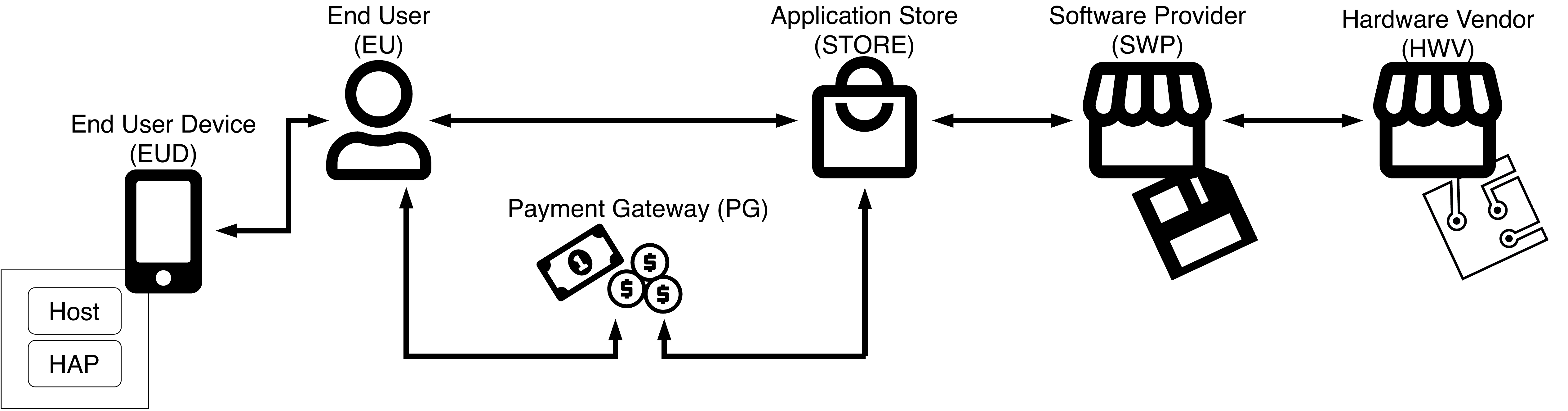}
	\caption{Actors involved in our scenarios showing a common mobile application deployment flow}
	\label{fig:actors}
\end{figure*}

We model the considered scenarios following a common scheme for application deployment for the mobile application market. The actors and their high-level interactions are shown in Fig.~\ref{fig:actors}.

We consider three main types of actors: 
\begin{enumerate*}[label=(\roman*)]
\item the software providers,
\item the hardware vendors, and 
\item the end users. 
\end{enumerate*}

The \SWPext (\SWP) is the entity that develops applications and sells them through several \emph{Application Stores} (\STORE) to reach a high number of potential customers. 

The \HWVext (\HWV) is the entity responsible for designing and selling the \HAPext (\HAP) used by the applications to load the \BS and it implements the soft IP cores developed and sold by \SWP{}s.
The \HAPext{}s considered in this paper are microprocessor-based Systems on Chips (SoCs) embedding state-of-the-art FPGAs featuring DPR and bitstream encryption mechanisms. 
The \HWV has full knowledge of the hardware it produces, including secrets and security mechanisms it embeds (\eg cryptographic keys). We assume that each hardware vendor has a publicly accessible service (\eg a web server) used to offer services to the software providers and to the end users (\eg prove hardware product authenticity, send firmware updates, etc.). It also provides publicly available information (\eg list of hardware capabilities, APIs, list of known compromised devices).
 
The \EUext (\EU) is the customer. The \EU may be interested in buying applications that have been developed by different \SWP{}s.
He/she owns an \EUDext (\eg smartphones, tablets, PDAs, etc.) that embeds the \HAP. The \EUDext and \HAP are two separate execution environments, therefore, they are considered as two separate entities in our scenarios. However, the \HAP does not directly access the network, all communications with the external world are mediated by the \EUDext. 

Together with the above three main actors, two more players are involved in the scenario we have considered in this paper. 
The \STORE acts as an interface between \EU{}s and \SWP{}s by collecting and making available to the \EU{}s software applications developed by different \SWP{}s. Every \STORE has a publicly accessible service (\eg a web server) and is connected to one or many payment gateways to allow customers to buy the software they want to purchase. Each \STORE knows the identity of its users.

Finally, the {\em \PGWext} (\PGW) offers payment services for software application purchases (e.g., using credit cards), interfacing with \SWP{}s and with \EU{}s.

\subsection{Security requirements}
\label{sec:assumptions:sec_req}

The high-level security objective of an \SWP{} is to preserve its intellectual property. Moreover, an \SWP aims at securing its applications, by preserving their authenticity, integrity and confidentiality.
The \EU{}s require to only execute authentic versions of the software they have bought.

We solely consider the security aspects of the hardware cores, while ensuring authenticity and integrity of executable code is out of the scope of this paper. Interested readers may refer to the literature on this field for a better understanding of available methodologies for software protection \cite{1027797,1212692,SDCSW5749894,SDCSW5727971}. 

The security requirements to be fulfilled in order to guarantee the authenticity and the intellectual property of all hardware cores deployed are the following ones:
\begin{itemize}

\item \emph{Bitstream confidentiality}: no other players but the \SWP itself must be able to read a bitstream in an intelligible form (e.g., if the application bitstream files are encrypted, the bitstream should not be available in plaintext). Every \SWP does not trust neither other \SWP{}s nor the Applications Stores used for distributing applications. Moreover, the \SWP does not trust the \EU{}s, even those who have legitimately bought its application. 

We have considered two scenarios in this paper. The first one (named Simple Scenario and presented in Section~\ref{sec:protocols:simple}) considers a case where a trust relationship exists between the \SWP and the \HWV{}s (\ie \HWV{}s might read the bitstream in plaintext) because of legal contracts or Non Disclosure Agreements (NDAs).
The second scenario, which tighten the confidentiality (named Full Scenario and presented in Section~\ref{sec:protocols:full}), relaxes this assumption, avoiding every trust relationship among all the parties involved.

\item \emph{Bitstream integrity and authenticity}: corrupted bitstream files must not be delivered to the \EU{}s, or used to configure the FPGA. The types of corruption to avoid are both accidental (\eg transmission errors) and deriving from intentional malicious alterations. Moreover, the \EU{}s must be ensured that the bitstream files they are loading in their \HAP are genuine, that is, they have actually been developed by the \SWP.
 
\item \emph{Legitimate \EUext}: only users who bought a legitimate copy of the software must be able to use the related bitstream files to configure the FPGA available in the HAP to assist the execution of the related software application. Nevertheless, in no case end users are allowed to access the bitstream in plaintext.

 %
\item \emph{Legitimate HAP}: only owners of legitimate \HAP can run applications into a user platform and load a \BS into the \FPGA. 
 

\item \emph{Need-to-know restrictions}: only the parties that actually need a sensitive information must have access to it. This is of great importance especially for information concerning the \EU{}s (e.g., the applications that they buy). This requirement is related to the privacy, however the need-to-know is subjective, thus it cannot be considered a privacy statement.

\end{itemize}
If any of these requirements is not met, the security of a bitstream is undermined and an adversary would be able to either read the bitstream and start reverse engineering or compromise the bitstream authenticity and integrity to inject malicious features, \eg by opening security backdoors.

\subsection{Attack model}
\label{sec:assumptions:attack_mdl}

We consider two type of attacks: IP attacks and integrity attacks. 

When a \emph{IP attack} is performed, an adversary tries to get access to an intelligible version of a bitstream. This attacks can be performed either on the \HAP by tampering with external memory devices or over the network links the application traverses during their deployment. This type of attack aims at violating the confidentiality of the bitstream in order to make illegal copies of the soft IP cores.
 
When an \emph{integrity attack} is performed, an adversary tries to compromise the integrity of a bitstream. This type of attack has two motivations: (1) a remote adversary may want to compromise the bitstream integrity to prevent the correct execution of the related software application (Denial of Service), (2) a malicious user may want to replace a bitstream file with a previous version (downgrade),  for example to avoid security updates.

In this paper we do not consider physical attacks that aim at damaging the \HAP or make it not operable (\ie hardware Denial Of Service attacks).

\subsection{Adversary model and security assumptions}
\label{sec:assumptions:adversaries}

The attacks just described can be carried out by a malicious attacker. 
We consider two types of adversaries trying to break the requirements introduced in Section \ref{sec:assumptions:sec_req}; remote and local adversaries.
We assume that both local and remote adversaries are knowledgeable attackers. 
They have considerable knowledge of the system and device they are attacking and they own high technical skills. 
Moreover, they have access to sophisticated tools and instruments \cite{Abraham:1991fk}.
Nonetheless, they are associated to different threat models.

A \emph{remote adversary} controls every possible network link involved in the proposed scenarios, \ie he/she can perform Man-In-The-Middle (MITM) attacks.
A Remote adversary uses techniques to intercept and understand the content of the messages exchanged between two arbitrary networking nodes. 
Once the attacker modifies the flow of messages, he can also make changes, delete and create completely new fake messages impersonating one of the communicating parties.

A \emph{local adversary} is a malicious \EU who has physical access to and full control of a \EUDext. That is, he/she can perform Man-At-The-End (MATE) attacks. 
Local adversaries have no restriction on the tools and techniques to reverse-engineer and then to tamper with the application (e.g., debuggers, emulators). Thus the application cannot be trusted to store/embed secret data or routines.
System libraries and general purpose libraries could be controlled by local adversaries, along with the operating system. In this case, to reach their goals attackers can use and alter system calls, the input/output subsystem, the network stack, the memory management subsystem and possibly others techniques.
Therefore, the communication with the \HAP mediated by software and drivers can be compromised by these adversaries, thus the data exchanged can be altered even if they are transmitted via secure channels that are safe against MITM attacks.
The attacker also controls the platform hardware. Every memory location can be read and written, including the processor registers. The attacker also controls the program storage medium, as a consequence he can read and change any of the stored bits at any time. This means that nothing can be considered secure in the user's environment.

The only part of the user's platform that we will consider secure is the \HAP, the stored information and the routines executed within the \HAP are considered confidential.
Indeed, we assume that, even if the \HAP may be placed in an hostile environment, the reconfigurable hardware and the security blocks surrounding it are resistant to physical attacks (\eg decamping the chip, side-channel attacks, etc.).

Finally, we assume that the local adversaries have no interest in performing Denial of Service attacks against their platforms and including \HAP (e.g., by repeatedly sending invalid bitstreams to block its functioning).


Competing firms are potential local adversaries, since they can invest non-negligible resources trying to compromise the security requirements.

Moreover, we assume that attackers all work under the {\em infeasibility} hypothesis. 
The infeasibility is associated to the derived computational cost, impossible to sustain for an attacker, in order to decipher the secret information needed by cryptographic algorithms or to invert digest algorithms. In practice, attackers are unable to solve exponential problems of proper size in useful time.

Eventually, when we state that two peers communicate with a $k$-secure channel, we indicate that the peers perform strong authentication and agree on a symmetric key $k$ that can then be used with data integrity and authentication algorithms and symmetric encryption algorithms to secure the exchanged data.



\section{Related works}
\label{sec:related-works}

Security concerns from the previously introduced model of software distribution are considered in this section.
A general overview of the feasible attacks against FPGAs, as well as a description of security issues and open problems regarding system security of FPGAs is provided by Wollinger et al. in \cite{Wollinger2004}. 
Security features, such as anti-tampering and data protection techniques, are described in \cite{6881481}. In \cite{7238102}, the authors define an FPGA threat model and evaluate how the security features offered by most FPGA vendors address the threats.

\subsection{Bitstream Confidentiality}
Guaranteeing bitstream confidentiality implies to protect from eavesdropping performed by external sources such as attackers or competitors. In this way, the information exchanged has to be maintained secure in the sense that it is not disclosed to unauthorized third parties. To maintain the confidentiality it is sufficient to encrypt the data.

Nowadays, bitstream confidentiality can be achieved resorting to the integrated bitstream encryption mechanisms offered by most FPGA manufacturers. 
These techniques allow to assist system designers in the protection of the secrecy of their IP cores, preventing the product to be cloned or reverse engineered.

The bitstream encryption operation is based on symmetric cryptography, \ie the key used to encrypt the information is the same used during the decryption.
The encryption key is generated by the system designer at design time and stored inside the FPGA.
The device decrypts the incoming bitstream during the configuration phase.
The FPGA stores the encryption key internally in a devoted memory, which could be backed-up by a battery (\eg BBRAM - Battery backed RAM) in order to maintain its content or a non-volatile one-time-programmable memory. The memory storing the key is designed to prevent physical attacks.
However, the encrypted bitstream configures the entire device. To decrypt partial bitstreams, system designers should build their own decryption engine requiring additional logic, thus reducing the usable area.  

Bitstream encryption is an effective solution to protect designer's IP against cloning or reverse engineering and IP disclosure \cite{Note:2008qf}.
However, in some cases encryption may not provide sufficient security, as reported in \cite{Moradi:2011, Moradi:2013}.
Another approach to protect IP cores against piracy and reverse engineering can be obtained through obfuscation. The authors of \cite{7857187} leverage FPGA dark silicon to obfuscate the functionality of the design.
Other works discuss techniques to consider for bitstream confidentiality as well as authentication\cite{drimer2007, 6482382}.

\subsection{Bitstream Integrity and Authenticity}
Since encryption alone does not protect the bitstream from modifications, integrity aims to ensure that the data exchanged does not undergo to modifications carried by unauthorized third parties across the network links during the transfer. 
Moreover, it provides assurance also from unintended modifications that might corrupt the data due to errors, \eg transmission errors.
If the integrity is maintained, also the system receiving the bitstream preserves its integrity.

Error control against data corruption can be accomplished with error-detection techniques, such as checksums and Cyclic Redundancy Checks (CRC).
To confirm that the configuration data stored in an FPGA device is correct, most vendors offer dedicated hardware for CRC features. 
However, CRCs are not suitable to protect against intentional malicious alteration of data \cite{Stigge:2006nx} since employ reversible functions.
Moreover, they are vulnerable to collision attacks, thus they do not guarantee adequate security levels. 

Hashing is another method commonly used to validate the integrity of information using a one-way function, \ie infeasible to invert. The fixed-length output of a cryptographic hash function, called message digest or simply hash, can be thought of as the fingerprint of the input bitstream, offering strong collision resistance. 
Cryptographic hashing primitives have been employed in \cite{Parelkar:2005qe, Parelkar:2005kl, drimer2007}.
Message digest, however, does not authenticate the source.
To verify the origin of the bitstream, the secret key shared between the vendor and the device can be combined together with a cryptographic hash function to generate a Hash-based Message Authentication Code (HMAC).
In \cite{Parelkar:2005kl} different authenticated encryption schemes have been evaluated and the dual-pass Counter with CBC-MAC (CCM) has been identified as the best option by lowest area footprint. However, Dual-pass authenticated encryption algorithms separate authentication and encryption procedures and therefore require significant overhead. 
Usually, authentication and confidentiality aspects are considered together, as proposed in \cite{drimer2007, Badrignans:2010qf}.

In presence of DPR, specific solutions to security problems, as well as safety \cite{SDCDPR6645549}, have also been discussed in literature.  
In \cite{Bossuet2004} a flexible security system based on bitstream encryption is proposed. While using FPGAs DPR, designers can freely choose encryption/decryption algorithms implemented as reconfigurable modules.
In \cite{Hori2008} authors developed a secure DPR system based on encryption of partial bit-streams with AES-GCM cipher. AES-GCM is an authenticated encryption cipher which guarantees both the confidentiality and the authenticity of a message.
In \cite{Kashyap:2016} is shown an improvement of the security of DPR in FPGAs re-encrypting a remotely received bitstream with a unique random key, while providing low area overhead and a high reconfiguration throughput.

\subsection{IP Licensing and Activation}
Providing confidentiality, integrity and authenticity of a bitstream protects the end-user device from malicious attackers. However, the intellectual property is left unshielded from piracy, which could damage the related business model of an application. IP theft introduces problems with design rights associated to the soft IP cores.
Encryption and obfuscation are strong tools to ensure security, but they cannot protect the IP in every stage of the life cycle.
Several works addresses these challenges \cite{6029984, 7987601, 7168561}. 
Enforcing IP security can be achieved resorting to solutions from a related area, \ie trusted computing. In particular, the Trusted Platform Module (TPM) is a microcontroller used to authenticate a target platform, enabling several cryptographic features \cite{TPM}. 
The TPM is embedded in and interacts with the target platform, providing cryptographic primitives for secure key generation and storage, random number generation and remote attestation. 
In literature, different methods of activation and licensing for IP cores protection have been employed: in \cite{4439246} a new scheme to track and control the licensed designs is presented, adopting public-key cryptograpy and symmetric encryption as well. 
The authors of \cite{7031902} propose an IP protection mechanism to restrict IP execution only on specific FPGA devices, limiting unauthorized copies and integration of the IPs. This solution, together with \cite{7987601}, enforces a pay-per-device licensing scheme for system developers, instead of purchasing unlimited IP licenses. 
Finally, a recent work \cite{7733105} provides a remote licensing and activation mechanism, guaranteeing anonymity for the end users.

\subsection{Open Challenges and contributions}
\label{sec:related-works:limitations}
The analyzed solutions bring several improvements related to the confidentiality, integrity and authenticity of a bitstream. 
However, the key employed for cryptographic operations is defined at design time and it is a secret shared between the system designer and the target device.
The system designer is therefore the only entity entrusted to provide new or updated bitstreams for a specific device. From this, it follows the limitation that software providers cannot generate bitstreams to assist their applications without relying onto the system designer. 
To the best of our knowledge, the available solutions are unable to fit the security requirements introduced in Section \ref{sec:assumptions}.


Our previous work proposed a simple architecture able to overcome this limitation. 
However, as will be described in Section \ref{sec:protocols:simple}, existing agreements between the software providers and the hardware vendors are necessary. The hardware vendors must be able to access the bitstream in plaintext, requiring a trust relationship with the respective software provider. This paper moves forward by presenting a new scenario, described in Section \ref{sec:protocols:full}, where the trust relationship among every of the involved parties is no longer required. Moreover, the users are able to gain additional privacy, limiting the identity disclosure to only software application stores.

In both cases, the confidentiality, integrity and authenticity of the bitstream exchanged (whether it is a new application or an update for an already-installed one) across the network links is maintained.

\section{Protocols and secure information exchange}
\label{sec:protocols}

This section introduces our solution for the secure reconfigurable computing model presented in Section \ref{sec:assumptions}. 
In particular, we will introduce and analyze the information exchange and the required protocols that will be used to identify the required hardware structures. 
Two cases will be analyzed: 
\begin{enumerate}
 \item a simple scenario fulfilling a reduced set of security requirements but exploiting minimum hardware facilities, and 
 \item a complex scenario featuring trusted computing hardware fulfilling the full set of security requirements of Section \ref{sec:assumptions}.
\end{enumerate}

In both scenarios, the target of the protocol is to define the operations to deploy a mobile application, composed of the software and its hardware counterpart, on the \EUDext. 
However, the transfer among the involved parties has to be secured with respect to the considerations introduced in Section \ref{sec:assumptions} and overcoming the limitations discussed in Section \ref{sec:related-works:limitations}.

\subsection{Case 1: Simple Scenario}
\label{sec:protocols:simple}
In this scenario, a simplified version of the protocol is presented.
This protocol satisfies a reduced set of security requirements, but benefits from simplicity and reduced hardware requirements.

\begin{figure*}[!htb]
	\centering



\def\hunit{1}
\def\vunit{1}
\def\hstep{3*\hunit}
\def\vstep{.85*\vunit}
\def\vmax{11*\vstep}
\def\labstep{1.27*\vstep}
\def\hblock{2.2cm}
\def\vblock{1cm}
\def\shifts{1.3}
\def\shiftw{1.3}
\def\shifth{1.5}
\def\gain{1.15}
\def\textmod{\small}

\def\id{id_{\mathrm{FPGA}}}
\def\hmac{\mathrm{H}}

\def\kcs{{K_{\mathrm{CS}}}}
\def\kss{{K_{\mathrm{SS}}}}
\def\ksh{{K_{\mathrm{SH}}}}
\def\kfpga{{K_{\mathrm{FPGA}}}}
\def\BS{\mathrm{BS}}
\def\SW{\mathrm{SW}}

\begin{tikzpicture}
\coordinate (x) at (\hunit,0);
\coordinate (y) at (0,\vunit);

\coordinate (FPGA) at (0,0);
\coordinate (FPGAEND) at ($(FPGA)-\vmax*(y)$);
\draw[black, thick] (FPGA) node[text width=\hblock,text badly centered, minimum height=\vblock, above,draw=black, thick,rectangle,inner sep=1pt] {HAP} -- (FPGAEND);


\coordinate (CLIENT) at ($(FPGA)+\hstep*(x)$);
\coordinate (CLIENTEND) at ($(CLIENT)-\vmax*(y)$);
\draw[black, thick] (CLIENT)  node[minimum height=\vblock,text width=\hblock,text badly centered,above,draw=black, thick,rectangle,inner sep=1pt]{Host} -- (CLIENTEND);

\coordinate (STORE) at ($(CLIENT)+\gain*\shifts*\hstep*(x)$);
\coordinate (STOREEND) at ($(STORE)-\vmax*(y)$);
\draw[black, thick] (STORE) node[minimum height=\vblock,text width=\hblock,text badly centered,above,draw=black, thick,rectangle,inner sep=1pt] {STORE} -- (STOREEND);

\coordinate (SW) at ($(STORE)+\shiftw*\hstep*(x)$);
\coordinate (SWEND) at ($(SW)-\vmax*(y)$);
\draw[black, thick] (SW) node[minimum height=\vblock,text width=\hblock,text badly centered,above,draw=black, thick,rectangle,inner sep=1pt] {SWP} -- (SWEND);

\coordinate (HW) at ($(SW)+\shifth*\hstep*(x)$);
\coordinate (HWEND) at ($(HW)-\vmax*(y)$);
\draw[black, thick] (HW) node[minimum height=\vblock,text width=\hblock,text badly centered,above,draw=black, thick,rectangle,inner sep=1pt] {HWV} -- (HWEND);


\coordinate (S1S) at ($(CLIENT)-1.3*\vstep*(y)$);
\coordinate (S1S2) at ($(CLIENT)-2*\vstep*(y)$);
\coordinate (S1E) at ($(S1S)+\gain*\shifts*\hstep*(x)$);
\coordinate (S1E2) at ($(S1S2)+\gain*\shifts*\hstep*(x)$);
\draw[->,thick] (S1S) -- (S1E) node[midway, above=-.1*\labstep] {\textmod
\begin{tabular}{c}
$\{r_C,\id,\mathcal{S},$\\
$\hmac(r_C,\id,\mathcal{S})\}_\kcs$
\end{tabular}};
 \draw[->,thick] (S1E2) -- (S1S2) node[midway, above=-.1*\labstep] {\textmod
\begin{tabular}{c}
$r_S,\mathcal{C},\hmac_\kcs(r_C,r_S,\mathcal{C})$
\end{tabular}};

\coordinate (S2S) at ($(S1E2)-.8*\vstep*(y)$);
\coordinate (S2E) at ($ (S2S)+\shiftw*\hstep*(x)$);
\coordinate (S2S2) at ($(S1E2)-1.5*\vstep*(y)$);
\coordinate (S2E2) at ($ (S2S2)+\shiftw*\hstep*(x)$);
\draw[->,thick] (S2S) -- (S2E) node[midway, above=-.1*\labstep] {\textmod
\begin{tabular}{c}
$\{r_{S2},\id,\mathcal{W},$\\
$\hmac(r_{S2},\id,\mathcal{W})\}_\kss$
\end{tabular}};
\draw[->,thick] (S2E2) -- (S2S2) node[midway, above=-.1*\labstep] {\textmod
\begin{tabular}{c}
$r_W,\mathcal{S},\hmac_\kss(r_{S2},r_{W},\mathcal{S})$
\end{tabular}};

\coordinate (S3S) at ($(S2E2)-.8*\vstep*(y)$);
\coordinate (S3E) at ($ (S3S)+\shifth*\hstep*(x)$);
\draw[->,thick] (S3S) -- (S3E) node[midway, above=-.1*\labstep] {\textmod
\begin{tabular}{c}
$\{r_{W2},\BS,\id,\mathcal{H},$\\
$\hmac(r_{W2},\BS,\id,\mathcal{H})\}_\ksh$
\end{tabular}};

\coordinate (S4S) at ($(S3E)-1*\vstep*(y)$);
\coordinate (S4E) at ($ (S4S)-\shifth*\hstep*(x)$);
\draw[->,thick] (S4S) -- (S4E) node[midway, above=-.1*\labstep] {\textmod
\begin{tabular}{c}
$r_H,\{\BS\}_\kfpga,\mathcal{W},$\\
$\hmac_\ksh(r_{W2},r_H,\{\BS\}_\kfpga,\mathcal{W})$
\end{tabular}};

\coordinate (S5S) at ($(S4E)-1*\vstep*(y)$);
\coordinate (S5E) at ($ (S5S)-\shiftw*\hstep*(x)$);
\coordinate (S5S2) at ($(S4E)-1.7*\vstep*(y)$);
\coordinate (S5E2) at ($ (S5S2)-\shiftw*\hstep*(x)$);
\draw[->,thick] (S5S) -- (S5E) node[midway, above=-.1*\labstep] {\textmod
\begin{tabular}{c}
$r_{W3},\{\BS\}_\kfpga,\mathcal{S},$\\
$\hmac_\kss(r_{W3},\{\BS\}_\kfpga,\mathcal{S})$
\end{tabular}};
\draw[->,thick] (S5E2) -- (S5S2) node[midway, above=-.1*\labstep] {\textmod
\begin{tabular}{c}
$r_{S3},\hmac_\kss(r_{S3},r_{W3},\mathcal{W})$
\end{tabular}};

\coordinate (S6S) at ($(S5E)-1.3*\vstep*(y)$);
\coordinate (S6E) at ($ (S6S)-\gain*\shifts*\hstep*(x)$);
\coordinate (S6S2) at ($(S5E)-2.3*\vstep*(y)$);
\coordinate (S6E2) at ($ (S6S2)-\gain*\shifts*\hstep*(x)$);
\draw[->,thick] (S6S) -- (S6E) node[midway, above=-.1*\labstep]{\textmod
\begin{tabular}{c}
$r_{S4},\{\BS\}_\kfpga,\mathcal{C},$\\
$\hmac_\kcs(r_{S4},\{\BS\}_\kfpga,\mathcal{C})$
\end{tabular}};
\draw[->,thick] (S6E2) -- (S6S2) node[midway, above=-.1*\labstep]{\textmod
\begin{tabular}{c}
$r_{C2},\{\BS\}_\kfpga,\mathcal{S},$\\
$\hmac_\kcs(r_{S4},r_{C2},\{\BS\}_\kfpga,\mathcal{S})$
\end{tabular}};

\coordinate (S9S) at ($(S6E2)-.3*\vstep*(y)-\hstep*(x)$);
\coordinate (S9E) at (S9S -| HW);
\draw[ thick,densely dashed,fill=black!20,nearly transparent] ($(S9S)+0.0*\vstep*(y)$) rectangle ($(S9E)-2*\vstep*(y)$);
\coordinate(X1) at ($(S9S)-.03*\hstep*(x)-.2*\vstep*(y)$);
\coordinate(X2) at ($(S9S)+.03*\hstep*(x)+.2*\vstep*(y)$);
\coordinate(X3) at ($(S9S)-.03*\hstep*(x)-.1*\vstep*(y)$);
\coordinate(X4) at ($(S9S)+.03*\hstep*(x)+.3*\vstep*(y)$);
\draw (X1) -- (X2);
\draw (X3) -- (X4);
\coordinate(X5) at ($(X1)+\hstep*(x)$);
\coordinate(X6) at ($(X2)+\hstep*(x)$);
\coordinate(X7) at ($(X3)+\hstep*(x)$);
\coordinate(X8) at ($(X4)+\hstep*(x)$);
\draw (X5) -- (X6);
\draw (X7) -- (X8);

\coordinate (S7S) at ($(S6E)-2*\vstep*(y)$);
\coordinate (S7E) at ($ (S7S)-\hstep*(x)$);
\draw[->,thick] (S7S) -- (S7E) node[midway, above=.5*\labstep] {\textmod$\mathrm{verify hash}$};

\draw[->,thick] (S7S) -- (S7E) node[midway, above=-.1*\labstep] {\textmod$\mathrm{load}(\{\BS\}_\kfpga)$};

\coordinate (S8S) at ($(S7E)-\vstep*(y)$);
\coordinate (S8E) at ($ (S8S)+\hstep*(x)$);
\draw[->,thick] (S8S) -- (S8E) node[midway, above=-.1*\labstep] {\textmod OK/KO};

%
%
%
%


\end{tikzpicture}
	\caption{The simple case workflow for downloading a new application. The grey area shows messages exchanged locally in the \EUDext, \ie between the host platform and the \HAP.}
	\label{fig:simpleworkflow}
\end{figure*}
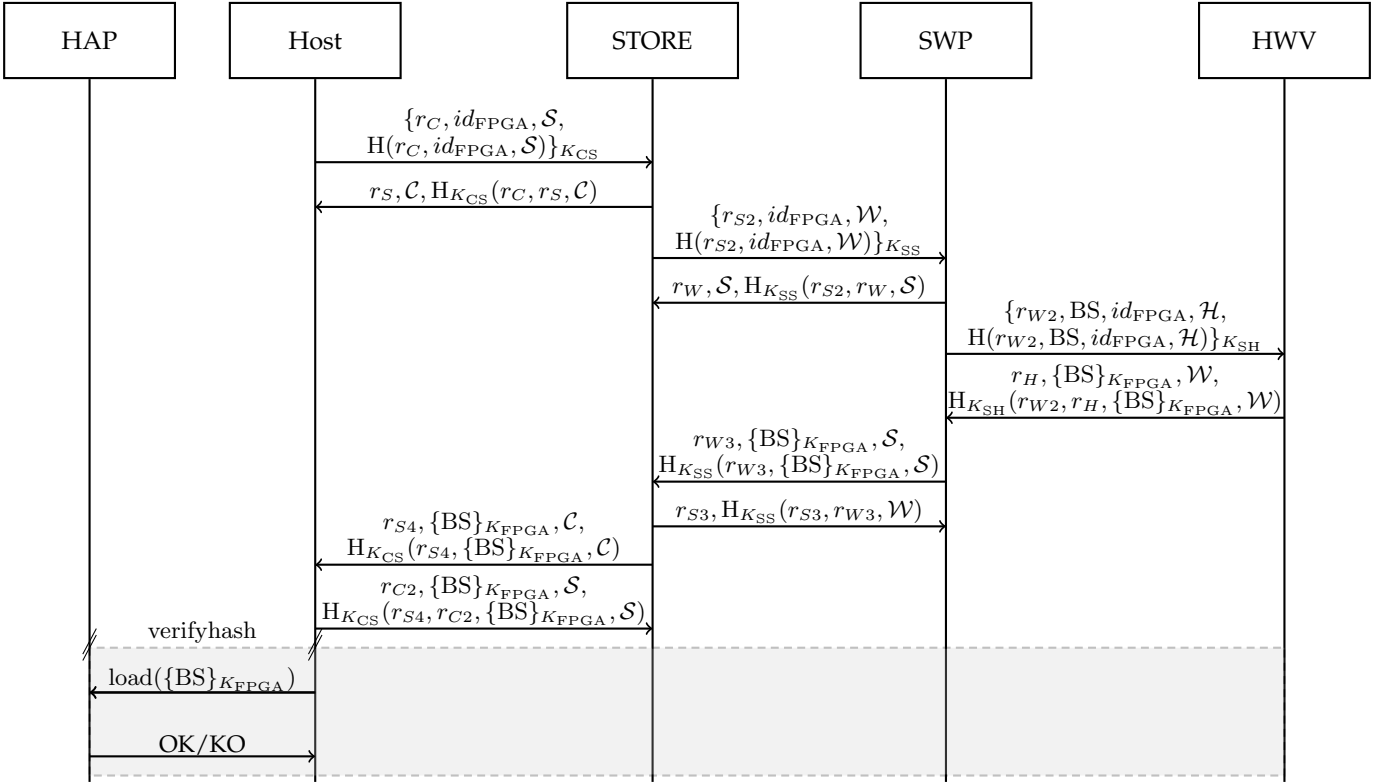

Fig.~\ref{fig:simpleworkflow} shows a possible workflow for the deployment and execution of an application onto a device embedding reconfigurable computing resources. 

In this simple scenario, we assume that there is a trust relationship between the \SWP and the \HWV. Moreover, together with the security requirements and assumptions presented in Section~\ref{sec:assumptions}, the following realistic functional assumptions are considered:
\begin{itemize}
	\item The \EU is able to access the store (\ie he has an account on the \STORE);
	\item The \EU has a credit card or any another equivalent payment system compatible with the store;
	\item The \STORE has existing agreements with one or more payment systems;
	\item The \HAP is identified by a unique code (\eg serial number), defined here as \IDHAP and stores a secret cryptographic key (\KHAP), both known by the \HWV. The key is not accessible from the outside of the \HAP;
	\item all involved entities are able to create secure channels (\eg through SSL/TLS protocols) satisfying confidentiality, data integrity and authentication, and peer authentication using an agreed symmetric key;
	\item the \SWP knows the \HAP manufacturers (\ie \HWV{}s) and can access to their services.
\end{itemize}

The following protocol presents the steps to be performed in order to buy an application and securely obtain the associated bitstream (see Fig.~\ref{fig:simpleworkflow}).

\begin{enumerate}
	\item ({\em The \EU buys the application}) The \EU browses the \STORE and decides to buy a mobile application, one that is composed by \SW and \BS, developed by a \SWP.
	As usual in the web market era, the \STORE redirects the user on the \PGW to complete the purchase. 
	The \PGW notifies the \STORE if the payment transaction terminates successfully. The \STORE starts the procedure to obtain the requested \BS to send to the client.
	The actual steps performed may change depending on the information exchanged between the \STORE and the \PGW, thus this is not reported in Fig.~\ref{fig:simpleworkflow}.
	
	\item ({\em The \EU sends the \STORE its data}) 
	The first interaction reported in Fig.~\ref{fig:simpleworkflow} involves the \EU and the \STORE. The \EU sends (from its device platform) the information needed to identify its \HAP, \ie \IDHAP.
	Since the communication involves sensitive data, the \EUDext and the \STORE use a secure channel that ensures confidentiality, data integrity and authentication by means of an agreed symmetric key $K_{CS}$\footnote{To simplify the presentation, we indicate that $K_{CS}$ is used both for symmetric encryption and to compute  a Message Authentication Code (MAC), even if $K_{CS}$ is better used as a master key to derive different keys (as actually done by TLS). 
	Also note that an authenticated encryption primitive could have been used instead of a MAC.}.
	The freshness is guaranteed by using a random number $r_C$. In Fig.~\ref{fig:simpleworkflow}, the symbols $r$ indicate random numbers.
	The \STORE then sends back an acknowledgement to the Client in the secure channel\footnote{The acknowledgments are sent to confirm the correct data receipt after all the interactions. Nonetheless, they are not explicitly reported in the text to ease the reading.}.

	\item ({\em The \STORE notifies the \SWP}) 
	The second interaction reports the \STORE that notifies the \SWP that a new customer bought the application (thus also its \BS). The \STORE forwards information about the \HAP of the client who bought the software. Again, the communication involves sensitive data (the \IDHAP) thus the \STORE and the \SWP communicate by using a secure channel that ensures confidentiality, data integrity and authentication by means of an agreed symmetric key $K_{SS}$.

	\item ({\em The \SWP sends the \BS to the \HWV}) 
	The \SWP sends the \BS of the purchased application and the \IDHAP that bought it. Also in this case, the communication involves sensitive data (\BS, \IDHAP) thus the \SWP and the \HWV communicate by using a secure channel that ensures confidentiality, data integrity and authentication by means of an agreed symmetric key $K_{SH}$.

	\item ({\em The \HWV prepares the \BS for the client's \HAP}) 
	After receiving the data, the \HWV ciphers the \BS with the \HAPkey and sends back to \SWP the ciphered bitstream \BSenc. In this case, the confidentiality of the secure channel is not needed, only the integrity and authentication. However, the already established secure channel could be used again.

	\item ({\em The \SWP forwards the encrypted \BS to the \STORE}) 
	the \SWP sends back to the \STORE the ciphered bitstream \BSenc via the available secure communication channel.

	\item ({\em The \BS ready to be downloaded by the \EU from the \STORE}) 
	Finally, the \STORE makes available to the \EU, \eg through the user account, both the executable code \SW and the ciphered bitstream \BSenc. 
\end{enumerate}

The \EU is then ready to install the application on its device.
After the installation, every time the user starts the application the reconfiguration of the \HAP will be triggered. Within the \HAP the \BS will be decrypted, thanks to the embedded controller that uses the \HAPkey \KHAP, and loaded onto the reconfigurable logic. 


It is worth noting that the proposed infrastructure can also be used to deliver updates to the \BS, while updates to the executable code can simply be downloaded by the \STORE through the usual methods.
When the \SWP updates the \BS, \eg because of a new version release or bug fixes, a reduced version of the previous protocol can be used: 
\begin{enumerate}
	\item the current version of the application initiates the communication with \SWP to check for available updates.
	
	\item If a new update is available, the client sends its \IDHAP to the \SWP (thus bypassing the \STORE) through a secure channel that ensures confidentiality, data integrity and authentication by means of an agreed symmetric key $K_{CS}$.
	The \SWP checks if the client corresponds to a valid customer\footnote{This check implies that the \SWP has stored the list of \IDHAP of customers sent by the \STORE.}.
	
	\item The \SWP connects to the \HWV via a secure channel and sends to the \HWV the \BS and the \IDHAP, encrypted with an agreed symmetric key $K_{SH}$;

	\item The \HWV ciphers the \BS using the \HAPkey \KHAP, and sends back to the \SWP the ciphered bitstream \BSenc. In this case, a secure channel is not needed however, the already established can be used;
	
	\item finally, the \SWP sends back to the \EU the updated application, which includes the updated ciphered bitstream \BSenc.
\end{enumerate}

Note that the simple scenario describes a communication protocol for the exchange of the bitstream among several parties that can be easily implemented with network nodes and a suitable \EUDext.


\subsection{Case 2: Advanced scenario}
\label{sec:protocols:full}

The simple scenario presented in Section~\ref{sec:protocols:simple} ensures confidentiality, data integrity and authentication among the involved parties. However, it requires a trust relationship between the \SWP and the \HWV, which might already exists due to legal contracts. Nonetheless, it may be a limitation.
Moreover, this protocol does not minimize the disseminated information. 
For instance, the \HWV is aware of each \BS application the \EU purchases.

%
%

In the simple scenario, the \EU is able to obtain an application for its device, while the \SWP achieves a secure transfer of its IP.
However, there is no assurance for the \SWP that its IP will not be loaded and executed in compromised hardware, which could ease the porting of attacks.
Additionally, the \EU privacy could be undermined (and the ``need to know'' requirement as well), since the other parties may collect information about the software bought by the user from any developer.
Instead, in the scenario presented here, either the \EU is able to preserve anonymity and the \SWP may only sell his application to users that own legitimate hardware. 
On the one hand, the full scenario drops the assumption that there exists a trust relationship between the \SWP and the \HWV. On the other hand, it requires that the \HAP offers more sophisticated features that are usually available at dedicated secure hardware. 
Indeed, the protocol implemented in the full scenario relies on the execution of the Direct Anonymous Attestation (DAA) protocol to guarantee all the security requirements in Section~\ref{sec:assumptions:sec_req}.

\subsubsection{The DAA Protocol}
\label{sec:protocols:full:daa}

The Direct Anonymous Attestation protocol has been designed to allow a verifier to check that a signature has been originated by a legitimate platform by means of a deterministic verification algorithm \cite{BrickellCC04}. 
Signatures can be used to convince a verifier about the integrity of the platform, that is, signatures allow achieving platform attestation. 
Signatures made with the DAA protocol are anonymous, that is, they do not leak any information about the signer, unless the signer wants a subset of signatures to be linked.
By using the DAA protocol it is also possible to detect the so-called \textit{rogue} platforms, so that the signatures obtained by means of compromised keys can be revoked. The manufacturer of the secure chips providing the DAA protocol can thus maintain a list of the compromised platforms.

The DAA protocol is independent of the public key authentication schemes, thus, different types of keys can be embedded in the secure hardware platform.

Different versions of the DAA protocol have been presented in the last years since the publication of the first version.
Some of them were provably secure under assumptions that do not guarantee the claimed security properties in the real world, for some other schemes, there exist known attacks to compromise them \cite{chen2008pairing,Brickell2009SSN,Chen10efficient,Chen2010efficient2}.
However, recently, two forms of DAA have been presented that have been formally proved against threats models that are not considered weak \cite{Camenisch20116SDH,Camenisch2016UCD}.

The DAA protocol has been standardized by the Trusted Computing Group (TCG) and is available in the Trusted Platform Module (TPM) since the version 1.2\footnote{However, the version of the DAA protocol implemented in the TPM v1.2 is no longer considered secure.}.
Later, in 2013, the TPM v2.0 has been developed and provided with the most efficient of the published DAA protocol versions.
The same version of the DAA protocol has also been inserted in the Intel processors as Enhanced Privacy ID (EPID) algorithm, which has been standardized as ISO/IEC 20008-1:2013\footnote{\url{https://www.iso.org/standard/57018.html}} and 20009-1:2013\footnote{\url{https://www.iso.org/standard/57079.html}}.

{
Three roles are involved in the execution of the DAA protocol:
\begin{enumerate}
	\item the \emph{DAA Issuer} is the entity that manufactures the secure hardware platform. The DAA Issuer knows all the secrets the secure hardware platform stores.
	\item The \emph{DAA Signer} is the secure hardware platform that produces the signatures used for attestation purposes. The DAA Signer entity is composed of two parts, the secure hardware platform and a Host, the set of all the software components that interfaces with the secure hardware platform (e.g., the computer system where the secure hardware platform is available, including the OS and all the drivers that allow accessing the hardware). Finally, 
	\item the \emph{DAA Verifier} is any external party (\eg a service provider) that is interested in verifying the integrity of the secure hardware platform.
\end{enumerate}

}


To achieve signature anonymity, the DAA protocol introduces two more features:

\begin{itemize}
\item a counter, which is a value used to generate multiple DAA keys from a single secret, and 
\item the basename: an optional value used to allow verifiers to link multiple DAA signatures signed under the same DAA key. Basenames can be considered pseudonyms and are obtained by means of a secure function computed on the secret of the secure hardware platform.
\end{itemize}

To abstract from the DAA protocol implementations (and their present and future flaws) and to make our result more general, we assume in this paper that the system uses the most secure and efficient DAA protocol version, which will expose the primitives presented below.
\begin{itemize}
	\item The DAA Setup is the procedure that a secure hardware platform performs to indicate its host and then to the DAA Issuer whether or not it is corrupted.
	\item The DAA Join is the procedure that a host performs, together with the secure hardware platform, to obtain the DAA Credentials (which can be seen as an anonymous public key certificate) and becomes part of the group of certified or attested secure hardware platforms.
	\item The DAA Sign/DAA Verify are the two procedures that the secure hardware platform and the host use to convince a verifier that the secure hardware platform is certified and the host has previously joined the group. These signatures are generated from a DAA Credential obtained after the join. This procedure is a form of remote attestation. Nevertheless, the same primitives could also be used to actually sign messages.
\end{itemize}

The DAA protocol ensures several security properties. They have been formally defined in previous works \cite{Camenisch20116SDH,Camenisch2016UCD} and are reported in this paper in an informal way:
\begin{itemize}
\item \textit{Completeness}: signatures from a valid basename of a honest platforms are accepted as valid by the verifiers.

\item \textit{Correctness of Link}: signatures created with the same basename by the same honest platform are correctly linked.

\item \textit{Unforgeability}: no adversary can create a signature that is recognized by the verifiers as a valid signature of a honest platform, regardless of the basename he used.

\item  \textit{Anonymity}: signatures of the same honest platform that use different basenames (or no basename at all) are not linked by any adversary. In other words, the adversary cannot tell if they are applied by the same honest platforms o by different honest platforms.
 
\item \textit{Non-frameability}: no adversary can create signatures with a given basename that links to a signature created by an honest platform. 

\end{itemize}

%

\subsubsection{Functional assumptions for the advanced scenario}

The following functional assumptions have been considered for the full scenario:
\begin{itemize}
	\item The \EU is able to access the store (\ie has an account on the \STORE);
	\item The \EU has a credit card or any another equivalent payment system accepted by the store;
	\item The \STORE has previous agreements with one or more payment systems;
	\item The \HAP owns valid DAA credentials and is able to run the Setup, Join, and Sign DAA methods;
\item The \SWP{}s know the \HAP manufacturers (\ie the \HWV{}s) and can access their DAA Issuer services;
	\item (optionally) the \STORE trusts the information obtained by the \HWV{}s through their services  (\eg the information about its compromised \HAP{}s).
\item (optionally) the involved entities are able to create secure channels (\eg through SSL/TLS protocols) satisfying confidentiality, data integrity and authentication, and peer authentication using an agreed symmetric key.
\end{itemize}

It is worth noting that the functional assumptions just mentioned are quite similar to the ones presented in \ref{sec:protocols:simple}. However, in this scenario, some entities offer special functionalities and play specific roles with respect to the context of the DAA protocol here employed:
\begin{itemize}
	\item The \HAP plays the role of the secure hardware platform;
	\item The \EUDext plays the role of the host where the secure hardware is attached;
	\item The \HWV plays the role of DAA Issuer (\ie the TPM manufacturer), thus it maintains the rogue oracle; 
	\item The \SWP plays the role of DAA Verifier (\ie it wants to authenticate the \HAP)
	\item optionally, also the \STORE can play the role of the verifier (\ie if it wants to authenticate the \HAP{}s)
\end{itemize}

\subsubsection{Workflow}

In the full scenario, before deploying an application onto a device embedding reconfigurable computing resources, the \HAP executes the DAA-Join protocol in order to be recognized as a trustworthy device by the \HWV. The target of this phase is to prove to the DAA Issuer that the secure hardware platform is trustworthy (setup) and to obtain valid DAA credentials (join). 

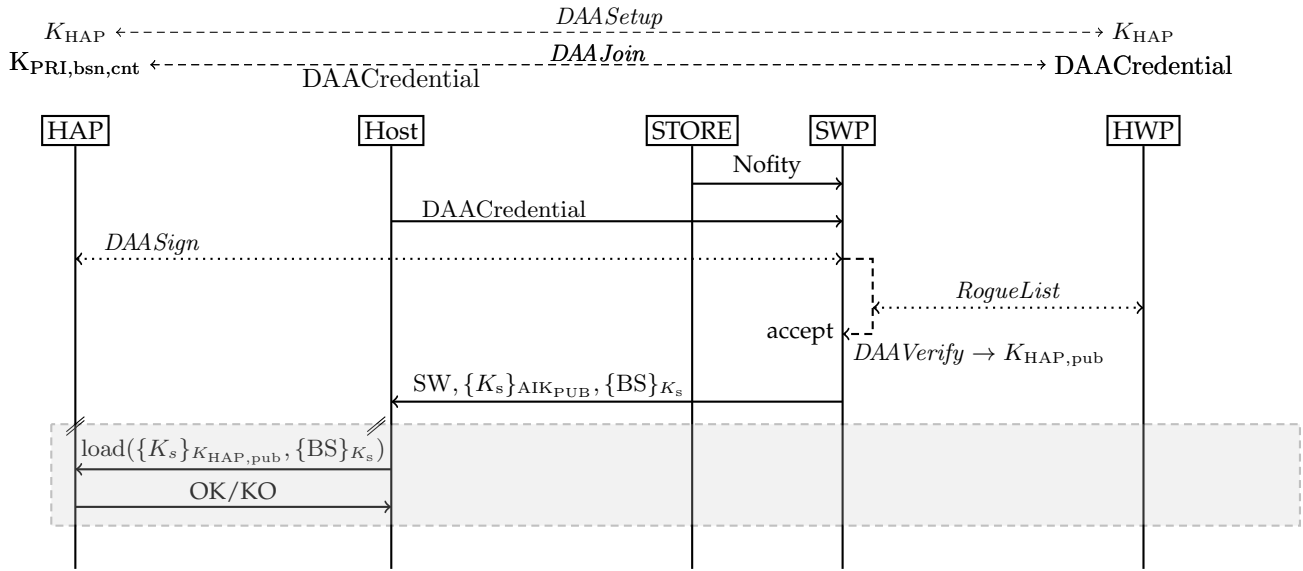
\begin{figure*}[!htb]
	\centering


\def\hunit{1}
\def\vunit{1}
\def\hstep{4*\hunit}
\def\vstep{.5*\vunit}
\def\vmax{11.25*\vstep}
\def\labstep{1.27*\vstep}

\def\textmod{\small}

\def\id{{\mathrm{DAA Credential}}}
\def\hmac{\mathcal{H}}

\def\kcs{{K_{\mathrm{CS}}}}
\def\kss{{K_{\mathrm{SS}}}}
\def\ksh{{K_{\mathrm{SH}}}}
\def\kfpga{{K_{\mathrm{HAP}}}}
\def\BS{\mathrm{BS}}
\def\SW{\mathrm{SW}}
\def\aik{{\mathrm{AIK_{PUB}}}}
\def\aikpri{{\mathrm{K_{PRI,bsn,cnt}}}}
\def\ks{{K_{\mathrm{s}}}}
\def\H{\mathrm{HAP}}
\def\gain{1.05}

\begin{tikzpicture}
\coordinate (x) at (\hunit,0);
\coordinate (y) at (0,\vunit);

\coordinate (FPGA) at (0,0);
\coordinate (FPGAEND) at ($(FPGA)-\vmax*(y)$);
\draw[black, thick] (FPGA) node[above,draw=black, thick,rectangle,inner sep=2pt, minimum height=22*\vstep] {HAP} -- (FPGAEND);


\coordinate (CLIENT) at ($(FPGA)+\gain*\hstep*(x)$);
\coordinate (CLIENTEND) at ($(CLIENT)-\vmax*(y)$);
\draw[black, thick] (CLIENT)  node[above,draw=black, thick,rectangle,inner sep=2pt, minimum height=22*\vstep ]{Host} -- (CLIENTEND);

\coordinate (STORE) at ($(CLIENT)+\hstep*(x)$);
\coordinate (STOREEND) at ($(STORE)-\vmax*(y)$);
\draw[black, thick] (STORE) node[above,draw=black, thick,rectangle,inner sep=2pt, minimum height=22*\vstep] {STORE} -- (STOREEND);

\coordinate (SW) at ($(STORE)+.5*\hstep*(x)$);
\coordinate (SWEND) at ($(SW)-\vmax*(y)$);
\draw[black, thick] (SW) node[above,draw=black, thick,rectangle,inner sep=2pt, minimum height=22*\vstep] {SWP} -- (SWEND);

\coordinate (HW) at ($(SW)+\hstep*(x)$);
\coordinate (HWEND) at ($(HW)-\vmax*(y)$);
\draw[black, thick] (HW) node[above,draw=black, thick,rectangle,inner sep=2pt, minimum height=22*\vstep] {HWP} -- (HWEND);



\coordinate (S1S) at ($(STORE)-1*\vstep*(y)$);
\coordinate (S1E) at ($(S1S)+.5*\hstep*(x)$);
\draw[->,thick] (S1S) -- (S1E) node[midway, above=-.1*\labstep] {\textmod Nofity};

\coordinate (S2S) at ($(CLIENT)-2*\vstep*(y)$);
\coordinate (S2E) at ($ (S2S)+1.5*\hstep*(x)$);
\draw[->,thick] (S2S) -- (S2E) node[near start, above=-.1*\labstep] {\textmod$\id$};

\coordinate (S3S) at ($(FPGA)-3*\vstep*(y)$);
\coordinate (S3E) at ($(SW)-3*\vstep*(y)$);
\draw[<->,dotted, thick] (S3S) -- (S3E) node[pos=.1, above=-.1*\labstep] {\textmod $\mathit{D\!A\!ASign}$};

\coordinate (S6S) at ($(S3E)$);
\coordinate (S6S1) at ($(S6S)+.1*\hstep*(x)$);
\coordinate (S6E) at ($(S3E)-2*\vstep*(y)$);
\coordinate (S6E1) at ($(S6E)+.1*\hstep*(x)$);
\coordinate (S6MID) at ($(S3E)+.1*\hstep*(x)-1.3*\vstep*(y)$);

\draw[->,densely dashed, thick] (S6S) -- (S6S1) -- (S6E1) -- (S6E) node[below right] {\textmod$\mathit{D\!A\!AVerify}\to K_{\H,\mathrm{pub}}$};
\node[left] at (S6E) {\textmod accept};

\coordinate (S4S) at ($(S6MID)$);
\coordinate (S4E) at (S6MID -| HW);
\draw[<->,dotted, thick] (S4S) -- (S4E) node[midway, above=-.1*\labstep] {\textmod$\mathit{RogueList}$};

\coordinate (S8S) at ($(S6E)-1.8*\vstep*(y)$);
\coordinate (S8E) at ($ (S8S)-.5*\hstep*(x)$);

\coordinate (S9S) at ($(S8E)$); 
\coordinate (S9E) at ($ (S9S)-\hstep*(x)$);
\draw[->,thick] (S8S) -- (S9E) node[pos=.65, above=-.1*\labstep] {\textmod$\SW,\{\ks\}_\aik ,\{\BS\}_\ks$};
\coordinate (S10S) at ($(S9E)-1.8*\vstep*(y)$);
\coordinate (S10E) at ($ (S10S)-\gain*\hstep*(x)$);
\draw[->,thick] (S10S) -- (S10E) node[midway, above=-.1*\labstep] {\textmod$\mathrm{load}(\{K_s\}_{K_{\H,\mathrm{pub}}},\{\BS\}_\ks)$};

\coordinate (S11S) at ($(S10E)-\vstep*(y)$);
\coordinate (S11E) at ($ (S11S)+\gain*\hstep*(x)$);
\draw[->,thick] (S11S) -- (S11E) node[midway, above=-.1*\labstep] {\textmod OK/KO};

%
%

\node(C4)[above =2*\labstep of FPGA] {\textmod$\kfpga$};
\node(S4)[above =2*\labstep of HW] {\textmod$\kfpga$};
\draw[<->, densely dashed] (C4) -- (S4) node[midway, above=-.1*\labstep]{ \textmod $\mathit{D\!A\!ASetup}$};

\node(C5)[above =1.25*\labstep of FPGA] {$\aikpri$};
\node(S5)at (C5 -| HW) {$\id$};
\draw[<->, densely dashed] (C5) -- (S5) node[midway, above=-.1*\labstep]{ \textmod $\mathit{D\!A\!AJoin}$};

\node(C5)[above =1.25*\labstep of FPGA] {$\aikpri$};
\node(S5)at (C5 -| HW) {$\id$};

\draw[<->, densely dashed] (C5) -- (S5) node[midway, above=-.1*\labstep]{ \textmod $\mathit{D\!A\!AJoin}$};

\node[below=-.1*\labstep] at (C5 -| CLIENT) {$\id$};

\coordinate(X1) at ($(S11S)-.03*\hstep*(x)+2*\vstep*(y)$);
\coordinate(X2) at ($(S11S)+.03*\hstep*(x)+2.4*\vstep*(y)$);
\coordinate(X3) at ($(S11S)-.03*\hstep*(x)+1.9*\vstep*(y)$);
\coordinate(X4) at ($(S11S)+.03*\hstep*(x)+2.3*\vstep*(y)$);
\draw (X1) -- (X2);
\draw (X3) -- (X4);
\coordinate(X5) at ($(X1)+\hstep*(x)$);
\coordinate(X6) at ($(X2)+\hstep*(x)$);
\coordinate(X7) at ($(X3)+\hstep*(x)$);
\coordinate(X8) at ($(X4)+\hstep*(x)$);
\draw (X5) -- (X6);
\draw (X7) -- (X8);
\draw[ thick,densely dashed,fill=black!20,nearly transparent]($(X1)+ .2 * \vstep *(y) -.05*\hstep*(x)$) rectangle ($(X1)+4.1*\hstep*(x)-2.5*\vstep*(y)$);

\end{tikzpicture}
	\caption{The full scenario workflow. The grey area shows messages exchanged locally in the \EUDext, \ie between the host platform and the \HAP.}
	\label{fig:fullworkflow}
\end{figure*}

Fig.~\ref{fig:fullworkflow} shows the workflow for the application deployment\footnote{Note that, to avoid making a too complex diagram, we have omitted all the authentication and integrity checks as well as the acknowledgements since they have been already presented in the simple scenario (and because they represent standard mutual authentication schemes that use random numbers \cite{menezes}).}.

The steps to buy a new application in the full scenario are the following ones (see Fig.~\ref{fig:fullworkflow}).
\begin{enumerate}
	\item ({\em The \EU buys the application})
	the \EU browses the \STORE and decides to buy an application, composed by a \SW and a \BS parts, developed by a \SWP. 
	The \STORE redirects the user on the \PGW to complete the application purchase. If the payment transaction terminates successfully, the \PGW informs the \STORE. The \STORE starts the procedure to send to the \EU the requested \BS.

	\item (The \STORE notifies \SWP of a new customer)
	the \STORE notifies the \SWP that a new customer intend to buy the application (thus he needs the related \BS). 
	Since the communication does not involves sensitive data, the \STORE and the \SWP may also avoid using a secure channel.
	
	\item ({\em The \EU sends the \SWP its data}) 
	During this interaction, the \EU sends the information needed to identify its \HAP, \ie the DAA Credentials issued by the \HWV (the DAA Issuer) when the DAA-Join was performed.
	Since the communication includes sensitive data (such as the credit card number for the purchase and the \IDHAP), the \EUDext and the \STORE will use a secure channel that ensures confidentiality, data integrity and authentication.

	\item (\HAP authentication and optional key exchange)
	the \SWP verifies that the customer owns a genuine \HAP by executing the DAA-Verify protocol. To know whether the \HAP is rogue, the \SWP connects to the \HWV services to query the ``rogue oracle''. If the verification fails, \ie the DAA Credentials correspond to a \HAP that is not genuine or they have been marked as rogue, the protocol is stopped by the \SWP. Some management policies will establish how to deal with these cases (e.g., ask installation on another platform or simply stop the purchase). 
	If the verification is successful then the payment is finalized.
	We assume that the interactions between the \HAP and the \SWP by means of the DAA protocol allow the exchange of a public key used by the \SWP to encrypt the data, which will be decrypted only inside the \HAP.
	In case the RSA-DAA protocol is used, based on the strong RSA assumption, the DAA Credential already conveys a certified public key $k_{\HAP,\mathrm{pub}}$.
	In case other public schemes are used, like the ones based on the qSDH and the LSRW assumptions or the ECDAA, we assume the \HAP is able to generate an RSA key pair and, via the DAA Sign, it can send the \SWP the signed public key $k_{\HAP,\mathrm{pub}}$.
	
	\item (The \SWP sends the \EU the encrypted bitstream)
	the \SWP generates a new symmetric key, the session key $K_s$, ciphers it using the received $k_{\HAP,\mathrm{pub}}$, then it ciphers the bitstream with $K_S$. Finally, it sends both the $\{K_s\}_{k_{\HAP,\mathrm{pub}}}$  and $\{BS\}_{K_s}$  to the \STORE\footnote{The data sent in this interaction can also be wrapped in a single enveloped data structure, e.g., the PKCS\#7 enveloped data.}.
	
	\item (The \EU downloads the application from the \STORE)
	finally, the \STORE makes available for download to \EU from his account both the software \SW, and the enveloped data structure including $\{K_s\}_{k_{\HAP,\mathrm{pub}}}$ and $\{BS\}_{K_s}$.  
	
\end{enumerate}

The \EU is then ready to install the application on its \EUDext.

As for the simple scenario, the \EU is then ready to install the application on its device.
Every time the \EUext starts the application, the \HAP is reconfigured with the soft IP core conveyed with the bitstream. 
Within the \HAP the \BS will be decrypted thanks to the embedded controller that deciphers the session key $k_s$ from $\{K_s\}_{K_{\HAP,\mathrm{pub}}}$ by using the corresponding key $K_{\HAP,\mathrm{pri}}$.

\section{Hardware architecture and implementation}
\label{sec:hwarch}
\begin{figure}[!htb]
	\centering
	\includegraphics[width=0.8\columnwidth]{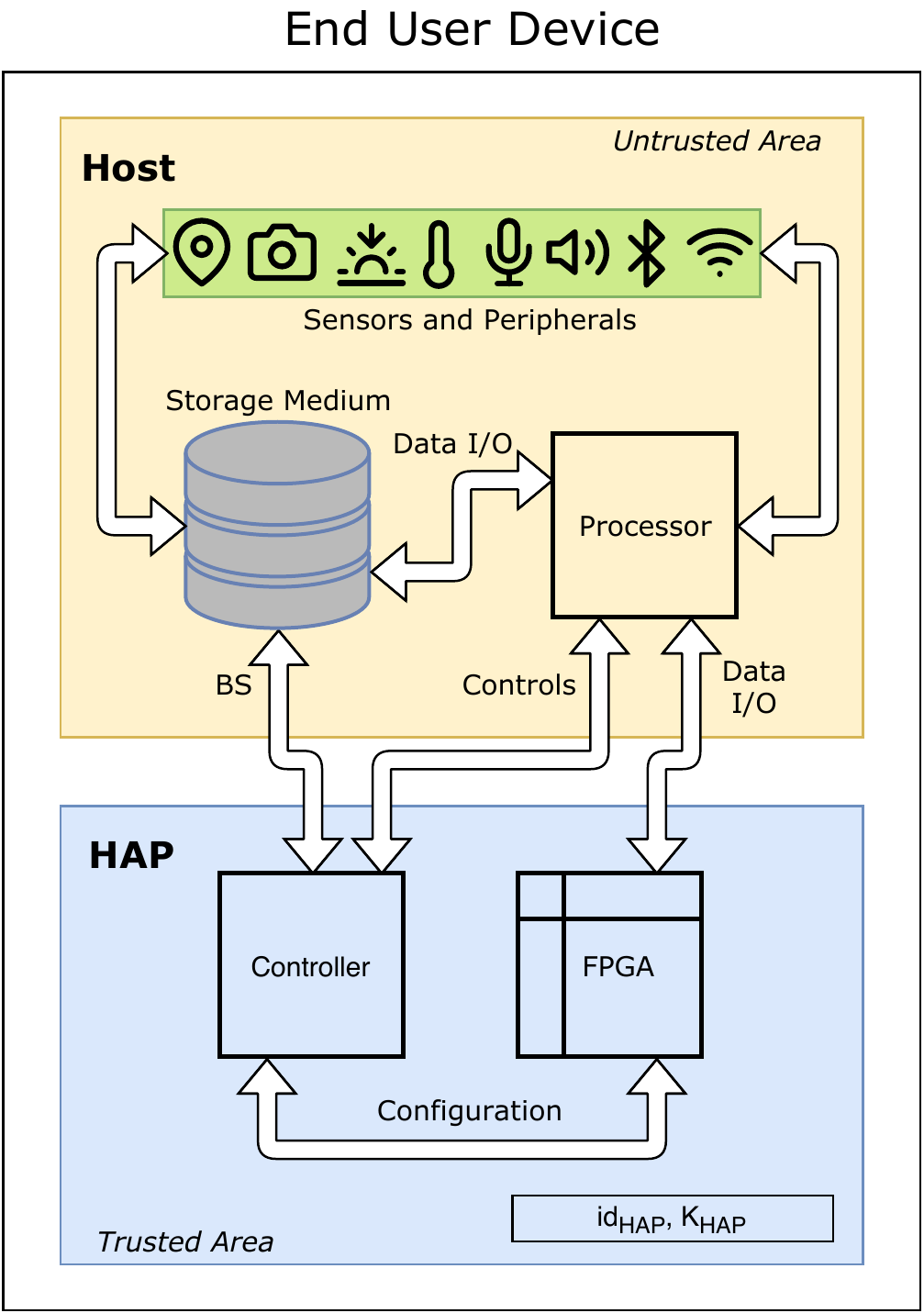}   
	\caption{The internal architecture of the End User Device}
	\label{fig:enduserdevice}
\end{figure}

In this section we present our proof-of-concept for the secure bitstream transfer protocol, detailing the proposed hardware architecture. 
We consider the \EUDext as a heterogeneous mobile device composed of the host platform and the \HAPext, as shown in Fig.~\ref{fig:enduserdevice}. 

\subsection{Architecture}
The host platform integrates hardware peripherals common for mobile computing devices, such as 
computational cores (\eg microprocessors, GPUs, etc.), sensors (\eg gyroscope, accelerometers, proximity sensors, light sensors, etc.), 
display or screen, memory and storage (\eg embedded memory, RAM, SD card, etc.) and network adapters for connectivity features.
The \HAP is embedded into the mobile device. Within the \HAP, there is the reconfigurable resource, \ie the FPGA. 
The \HAP embeds also a microcontroller offering cryptographic functionalities, such as key generations, and to securely load the bitstreams of the applications installed onto the FPGA.

\subsection{Interconnections}   
The class of mobile devices is often characterized with components enabling two different types of connections, \ie wired and wireless.
Network adapters and antenna of the \EUDext enable external connectivity towards an internet connection, which is necessary for the user to browse application stores and to download the \SW and \BS of the applications to be installed.   

Internal connections depend on the actual hardware architecture of the \EUDext. Among the components of the \EU device, the data path of the bitstream spans from the storage medium to the FPGA within the \HAP.   
In order to permanently store an application, the host microcontroller is connected to the storage medium.   
The hardware description of applications is loaded onto the FPGA with a link between the host and the \HAP.   
Eventually, when the FPGA has to be programmed the bitstream is moved to the \HAP where the controller decrypts the hardware configuration and sends it to the FPGA.   

\subsection{Implementation details}
To simulate our architecture, we separated the host from the \HAP of the \EUDext.
The host device is emulated with a normal laptop/desktop PC connected to the internet, running the client software. The \HAP is connected to the host PC with a USB cable. For prototyping the \HAP, we employed a particular chip, \ie SEcube\texttrademark~\cite{secubeds}. SEcube\texttrademark{} is a heterogeneous system-on-chip, embedding three components interconnected within the same package: a microcontroller, an FPGA and a SmartCard.
The microcontroller is a 32-bit low-power ARM Cortex-M4 processor performing the necessary operations to communicate with the host, through USB and SDIO interface. To securely program the FPGA, the microcontroller is connected to the FPGA through a 16-bit wide bus. The reconfigurable hardware is a Lattice MachXO2-7000 low-power FPGA. The SmartCard is Certified Common Criteria CC EAL5+.
The host is connected to the storage medium, \ie a microSD card, which is accessible also from the microcontroller of the \HAP. The storage medium is employed to store the bitstream of the applications downloaded. Although the \BS on the storage medium can be accesible from outside, it is stored encrypted to avoid confidentiality breaches. The bitstream is decrypted only within the \HAP, which is considered a trusted area.
The other involved parties \STORE, \SWP and \HWV are represented with other PCs connected in a network. Each entity executes an application to communicate with the other network nodes and serves the respective clients according to the protocols disccussed in Section~\ref{sec:protocols}.

\subsection{Software stack}
The software executed on the bare metal of the \HAP device is the open source firmware of the SEcube\texttrademark{} SDK. 
It provides a high-security abstraction layer through API functions, ranging from encryption and decryption utilities to cryptographic keys management for the keys embedded in the \HAP, as well as interfaces for secure communication with the host platform.
To implement the functionalities required for the DAA protocol of Section~\ref{sec:protocols:full}, we adopted the MIRACL Crypto SDK~\cite{miracllib}. The MIRACL SDK is a C/C++ library providing optimized implementations for security primitives (\eg elliptic curve cryptography - ECC), tailored for constrained environments, such as embedded systems and mobile devices.

Among the applications running on the host side of the \EUDext, the counterpart of the APIs are used to communicate with the \HAP. 
Also, the host device runs the client market application, which is able to connect to the services offered by the \STOREext{}s. 
The server applications of the \STORE, \SWP and \HWV resides in the same PC, working as separate asynchronous processes.
The services of these entities have been emulated as Python3 applications. For the client/server architecture and asynchronous communication functionalities we employed the \emph{asyncio} module. The communication among these processes, running on the same physical machine, flows through different port numbers of the communication sockets on the same IP address. Every message sent is encrypted and signed to guarantee integrity, confidentiality and authenticity. When reaching the destination, the signature of the message is checked against malicious alterations. 
When the download of the bitstream is completed, the application running on the \EUDext triggers the mechanism for loading the bitstream onto the FPGA. The \HAP firmware retrieves the encrypted data from the storage medium and check its integrity and authenticity. Finally, the bitstream is decrypted and the plaintext configuration is loaded onto the FPGA by the microcontroller equipped onto the \HAP.

\section{Security analysis}
\label{sec:security_analysis}

From Section \ref{sec:assumptions:attack_mdl}, we see two types of attackers to counteract: man-in-the-middle (MITM) and man-at-the-end (MATE) attackers.
The implicit assumption in an MITM scenario is that both the endpoints are trusted entities. However, this assumption is no longer valid in the case where also MATE attackers are interested in obtaining the bitstream. Indeed, MATE attacks are more challenging to prevent.
However, all the security relevant operations are performed in the \HAP, which we assume is a trusted device that cannot be attacked by our adversaries.

It is worth remarking that the proofs of the security of these solutions hold under the {\em infeasibility} hypothesis. 
That is, we assume the use of state-of-the-art secure cryptographic algorithms with proper key lengths.
For instance, currently, a 256-bit security is required from symmetric encryption, that is, the best attack should be a brute force attack on a $2^{256}$ size key space, \ie using AES256 guarantees an appropriate level of security. Furthermore, at least 80-bit of security is required for the digest algorithms, for instance SHA-256 is currently a valid choice.

Moreover, we recall that when two communicating peers $A$ and $B$ share a symmetric key $k$:
\begin{itemize}
  \item (\textit{confidentiality}) only $A$ and $B$ are able decrypt messages encrypted with $k$;
  
  \item  (\textit{symmetric integrity and authentication}) if $A$ receives a message and a MAC computed by using $k$, $A$ deducts that the message and the MAC were generated by $B$ and no one changed the original message, analogous considerations are valid when $B$ receives messages and MACs $A$.

\end{itemize}


\subsection{Simple Scenario}
\label{sec:security_analysis_simple}

MITM attacks performed by remote adversaries represent a standard security problem in computer networks and there are provably secure solutions to protect against these attacks: the channel protection techniques.
In fact, MITM attacks can be neutralized by using strong peer authentication mechanisms to avoid impersonation, symmetric data integrity and authentication techniques to avoid the message forging and alteration, and symmetric data encryption to ensure confidentiality of exchanged data. 

The techniques we adopted in the simple scenario protocol ensure data confidentiality by using symmetric encryption algorithms (\eg AES), and symmetric data integrity and authentication algorithms (\ie a keyed digest or HMAC using a cryptographic hash function) used in the a ``challenge-response (keyed) one-way functions authentication protocol'' \cite{menezes}.
These are well-known approaches that are provably secure under the infeasibility assumption.

The first step of the protocol for bitstream IP protection specified in Section \ref{sec:protocols:simple} represents a typical e-commerce scenario very widespread nowadays, where a user is assumed to have a credit card, a TLS-enabled browser installed in his environment, and an account on the application store of the platform. The store relies on payment services that should adhere to the Payment Card Industry Data Security Standard (PCI DSS)  to work in the financial world \cite{PCIDSS}.
Indeed, the PCI DSS imposes high security requirements for merchants and payment servers that store, process or transmit payment cardholder data when implementing a robust payment card data security process.

By analysing the protocol, we derive that the \IDHAP is only readable by the \EU, the \STORE, the \SWP, and the \HWV.
In fact, the \HAP identifier is encrypted with the key shared between the client browser and the \STORE, then the \STORE encrypts it with the key shared with the \SWP. Finally, \IDHAP is again encrypted with the key shared with \HWV and it is sent to the \HWV.
MITM attackers cannot read the \HAP identifier if strong encryption algorithms are used. 
Additionally, impersonation attacks are impossible as the sent messages allow symmetric authentication of the party.
Furthermore, the data authentication and integrity mechanisms used by the presented protocol (\ie HMAC or an authenticated encryption algorithm) prevent that modifications to exchanged messages are not detected. Therefore, the only remaining attacks are the DoS attacks, which we have excluded as it is not among our goals (and not easy to prevent in all cases).

Even more important, the \BS is read in clear only by the \SWP and the \HWV. In fact, the bitstream is encrypted with the key shared between the \SWP and the \HWV. Then, the \HWV encrypts the \BS by using the secret key $K_{\HAP}$, shared with the \HAP of the End User Device. Therefore, no one but the \HAP with the identifier exchanged during the protocol is able to load and use it. 
The received bitstream is stored in an encrypted form until it is moved to the \HAP decrypted and loaded onto the \FPGA, while the software application is usually downloaded in plaintext, since other software protection techniques are displaced to protect the intellectual property, if needed.
Therefore, MITM and MATE attacks that aim at reading the bitstream in intelligible way are avoided.

These considerations prove that only the \EU{}s that bought the software are able to use the corresponding \BS. Moreover, even the end users are not able to read the plaintext of the bitstream.

Note that the confidentiality of \IDHAP and the integrity and authentication of all messages could be achieved if, instead of an ad-hoc protocol as in Fig.~\ref{fig:simpleworkflow}, general purpose channel protection mechanisms are used. 
Even better, these channels usually also provide protection from reply attacks and filtering, by numbering exchanged packets or by storing the last packets.
The most widespread ones are the TLS protocol \cite{rfc5246}, which works at the transport layer of the ISO/OSI stack, or the IPsec protocol \cite{ipsec}, which works at network layer, and other application layer methods, usually message protection techniques (\eg WS-Security).
In our case the TLS approach is the preferred one, which better integrates with a web-based scenario that is very frequent in the e-commerce scenarios.
Indeed, it does not require additional software or any previous knowledge of the other communicating party but limited modifications of the services or the availability of an ad hoc API. 

On the other hand, there is no alternative than the explicit encryption with $K_{\FPGA}$ to protect the confidentiality of the bitstream.

\subsection{Advanced Scenario}
\label{sec:security_analysis_full}


The advanced scenario falls in the same common e-commerce use case. We propose an alternative use of DAA protocol in a more general context than the TCG related ones.
The only difference is that the user is required to perform a DAA\_Setup to initialize the \HAP and a DAA\_Join before running the protocol to generate and issue new DAA credentials of the \FPGA.
From the functional point of view, DAA-related operations are only available in experimental settings and have not yet reached a large audience, thus not to current customers. However, they do not pose significant challenges for future \EU{}s, also because they can be mostly automated and their complexity can be properly hidden to end users.

Indeed, after buying the software, the client is redirected towards the \SWP and he is asked to perform a DAA\_Sign to attest the integrity of its \HAP. 
Only the \SWP knows the session key $K_s$, which is then ciphered with the public key received by \HAP.
Therefore, $K_s$ will only be available inside the \HAP, in the trusted part of the protocol entities. This in turn allows satisfying the  ``Bitstream confidentiality'', ``Bitstream integrity and authenticity'' and ``Legitimate End User'' requirements.
To complete the integrity verification the \SWP will contact the ``rogue oracle'', that is, the entity knowing the IDs of the compromised \HAP. The \HWV is the best (and only, to date) candidate to maintain this list.

Note that the generation of an RSA key pair whose public component has to be signed with the DAA Credentials and then sent \SWPext, is currently not implemented in the TPM chips.
However, it is not against the TPM and DAA specifications, as DAA\_Sign can be both used for remote attestation and data signature purposes.
Therefore, from the functional point of view, the operation we require poses integration issues with current TPM chips but it is not a problem for possible future versions of the TPM nor for the secure hardware (FPGA) we are considering for this paper, as they are fully reconfigurable.

Therefore, in this scenario the only entity that knows the \BS is the \SWP, because the \HAP can only load and use it internally. 
Thus, when following  this approach the ``Need to know'' principle of the \BS is better ensured as it guarantees that the minimum number of entities know the \BS in an intelligible form.
Indeed, as the \HAP of the \EUext is the only entity that knows the private component of the public key sent to the \SWP, only the \HAP of the legitimate buyer can use the \BS, thus ensuring the ``Legitimate \EUext''.

Additionally, by using the DAA\_Sign and DAA\_Verify operations and the remote verification using the rogue oracle, the \SWP is able to sell its products to buyers whose \HAP is not compromised, thus proving the ``Legitimate \HAP'' requirement.

Therefore, we can conclude that the security of the advanced scenario only depends on the correctness and security of the DAA protocol. We assume that the DAA protocol used is correct and secure. Given the level of interest in this protocol and the effort put by the TCG and the researchers in the field, this is currently a reasonable assumption and it will become even more acceptable also in the near future. 
Indeed, as discussed in Section~\ref{sec:protocols:full:daa}, the recent progresses and publications give positive hints on the fact that the DAA development is converging to a form that is both secure and not too demanding from the computational point of view \cite{Camenisch20116SDH,Camenisch2016UCD}.

The idea of using the DAA protocol has major consequences to the impact of the work presented here. 
Indeed, the use of a well-known protocol guarantees a bigger control on the strength of the secure mechanisms. 
Moreover, there will be more people researching for flaws, as the impact of attacks becomes very high thus also the impact of publication of such attacks is more likely to reach a higher visibility.
Also reactions of the involved parties to potential flaws would have much more support.
The TCG is also increasingly working to increase anonimity and improve users's privacy. In our opinion this is a major trend that is worth joining.

We reach at the same time a strong protocol that guarantees a better level of user privacy (\ie used data and hardware identifiers) and minor exposition of the company IP (soft IP core are only read by developers and used by secure hardware).

Therefore, even if we do not necessarily propose the use of TCG-specified TPM chips, we follow the progresses in this field and re-implement in reconfigurable hardware the same features.



Another good reason to join a mainstream initiative, is a greater level of control on the whole supply chain of the trusted hardware device, being they TPMs of FPGA-based devices. Indeed, since nothing can be done if the \HWV inserts backdoors in the produced devices, we have to resort to a trusted supply chain.

Together with simple backdoors to access the content of the \HAP to download the IP cores that need to be protected, other attacks against the whole End User Platform can be leveraged by holes in the security of the supposedly trusted device.

As a simple instance of potential attacks, instead of generating a new RSA key pair inside the \HAP, the \HWV can generate and inject a certain number of RSA key pairs to choose from (or DAA Credentials if RSA-DAA is used). Therefore, the \HWV would have access to all bitstreams encrypted with those public keys.


\subsection{Physical Attacks}

Physical attacks represent another type of possible attacks performed directly on the \EUDext.
In this case, an attacker aiming at finding the symmetric encryption key must be necessarily a MATE owning the device of interest. We distinguish the case of non-invasive physical attacks from invasive physical attacks. 

Non-invasive attacks can be partitioned in attacks where the device is purposely \emph{stressed}, \ie \emph{active}, and attacks where there is no interaction with it, \ie \emph{passive}.
In both cases, the destruction of the device is not necessary. However they generally require a longer time to be accomplished.
Brute-force attacks have already been excluded under the infeasibility hypothesis. 
Side-channel attacks can still be employed, however they are not always possible and require specialized equipment.
The SEcube\texttrademark{} chip adopted for the prototype is secure against the differential power analysis attack as shown in \cite{8110067}.
Timing attacks can be neutralized resorting to constant-time implementation of the bitstream decryption and other security primitives.

Invasive physical attacks destroy the device and require sophisticated and expensive equipment, knowledgeable attackers, and possibly long time to be carried out.

In our scenarios, a physical attack might target the Host platform, the \HAP or the storage medium.
Any attack to the Host or the storage medium is neutralized with the encryption algorithm, since these parts of the end user platform are considered not trusted. Accessing these peripheral will give to the attacker the possibility to find the ciphertext of the bitstream. But the data must still be deciphered, by reversing the key.
The \HAP is in fact the critical element, since the bitstream is deciphered within this device through the key stored here.
Local adversaries should first bypass any external protection to reach the internal circuitry. 

If we suppose that an invasive physical attack is successful, only a single device is compromised. Every device stores a unique serial number and the cryptographic key relative to the specific device. In this way, an attacker wanting to inject vulnerabilities or malicious hardware must repeat the attack on other \HAP{}s. This means that also other devices should be compromised in the same way, leading to higher costs to spread the attack. Also, the effort to exploit the attack is linear with the number of device to compromise.

Nonetheless, if the attack is successful, the target IP stored onto the device is compromised by only compromising a single device. 
Indeed, when the bitstream is available to the attacker, it can be decrypted and its description might not remain confidential, but could be disclosed to the public. Moreover, this security breach gives the possibility to the attacker to recover also the \IDHAP of the compromised device. By knowing this information, an attacker could also obtain any other bitstream previously bought for that specific device.

Note that if a TPM chip supports the DAA protocol features, this chip is protected by design against side-channel (timing information, power consumption, electromagnetic leaks) and physical attacks. Additionally, as it uses cryptographic operations it must comply the FIPS 140-2 standard.

The TPMs available on the market provide memory curtaining and protected execution to avoid that the MATE reads the stored secrets. Therefore, the only way to read the secrets is physically tampering with the chip. To the best of authors' knowledge, the only successful physical attack is the one presented at the Black Hat conference 2010 \cite{blackhat2010}. At cost of six months and 200,000\$, Tarnovsky tampered the internal circuitry of an Infineon TPM of the SLE 66PE family of contactless interface microcontrollers to get the secrets. The attack required to dissolve the outer shell with chemicals and remove the layers of mesh wiring to access the chip's bus to read the secrets by tapping the communications channels using small needles.
The attack was defined by the author ``not easy to duplicate'' and the Trusted Computing Group that issued the TPM specifications, a bit more optimistically, ``exceedingly difficult to replicate in a real-world environment''.

Some other attacks to the TPM are known in literature that are not important for us.
The reset attack, is a method to reset the TPM without resetting the entire system. In this way the known-good hashes can be stored in the TPM circumventing the ``extend-only'' functionality that does not allow to overwrite values in the TPM registers. This attack is mounted against TPM family 1.1 using a vulnerability of the Low Pin Count (LPC) bus that allows to reset all the attached devices. From the family 1.2 this attack is no longer possible. 


Another reset attack is presented in \cite{217652}, however the authors already contributed together with manufactures with a patch to solve the vulnerabilities.

A recent side channel attack on TPM 2.0 devices is presented in \cite{244048}, where the authors are able to recover 256-bit private keys for ECDSA signatures Intel firmware-based TPM as well as a hardware TPM. The attack exploits timing information leakage and allows key recovery in less than two minutes. Also in these case, the vulnerabilities have been solved.

\section{Conclusion}
\label{sec:conclusion}

This work addresses the protection of soft IP cores to be deployed in the context of mobile heterogeneous systems.

We provide an architecture for a secure transfer of a soft IP core bitstream from a generic software developer to an end user owning a device equipped with reconfigurable logic, 
considering a common real-world scenario of mobile application market.

The provided protocol details the deployment for two scenarios, considering first a minimum set of security requirements, then an advanced scenario fulfilling tighter security properties.

We guarante that only legitimate users purchasing a legitimate copy of an application are enabled to use it. 
During the deployment, we are able to maintain the confidentiality of the intellectual property. Also, the integrity of the data transferred is preserved to guarantee that no alteration, whether malicious or not, has been performed.
Compliance to these requirements protects from MITM attackers. We considered the threat of MATE attackers as well. 
Finally we also provided a prototype implementation of the whole architecture, employing a system-on-chip as heterogeneous system to work together with a host device (\eg a PC, or a mobile device).

\ifCLASSOPTIONcaptionsoff
  \newpage
\fi




\bibliographystyle{IEEEtran}

\begin{thebibliography}{10}
\providecommand{\url}[1]{#1}
\csname url@samestyle\endcsname
\providecommand{\newblock}{\relax}
\providecommand{\bibinfo}[2]{#2}
\providecommand{\BIBentrySTDinterwordspacing}{\spaceskip=0pt\relax}
\providecommand{\BIBentryALTinterwordstretchfactor}{4}
\providecommand{\BIBentryALTinterwordspacing}{\spaceskip=\fontdimen2\font plus
\BIBentryALTinterwordstretchfactor\fontdimen3\font minus
  \fontdimen4\font\relax}
\providecommand{\BIBforeignlanguage}[2]{{%
\expandafter\ifx\csname l@#1\endcsname\relax
\typeout{** WARNING: IEEEtran.bst: No hyphenation pattern has been}%
\typeout{** loaded for the language `#1'. Using the pattern for}%
\typeout{** the default language instead.}%
\else
\language=\csname l@#1\endcsname
\fi
#2}}
\providecommand{\BIBdecl}{\relax}
\BIBdecl

\bibitem{Albright:fk}
\BIBentryALTinterwordspacing
P.~Albright. {Become a Mobile Apps Innovator: Picking an {OS} and learning to
  monetize are key}. [Online]. Available:
  \url{http://www.computer.org/portal/web/buildyourcareer/HS26}
\BIBentrySTDinterwordspacing

\bibitem{appstorelist}
A.~Dogtiev. (2019 (accessed July 8, 2019)) App stores list.
  \url{https://www.businessofapps.com/guide/app-stores-list}.

\bibitem{iresearch}
\BIBentryALTinterwordspacing
iResearch. {Worldwide mobile app revenues in 2014 to 2023}. [Online].
  Available:
  \url{https://www.statista.com/statistics/269025/worldwide-mobile-app-revenue-forecast/}
\BIBentrySTDinterwordspacing

\bibitem{Ke-fei:2010ly}
S.~Ke-fei, ``{Application of FPGA in Aerospace Remote Sensing Systems},''
  \emph{OME Information}, 2010.

\bibitem{Surratt:2005ve}
M.~Surratt, H.~Loomis, A.~Ross, and R.~Duren, ``{Challenges of remote FPGA
  configuration for space applications},'' in \emph{{Aerospace Conference, 2005
  IEEE}}.\hskip 1em plus 0.5em minus 0.4em\relax Ieee, 2005, pp. 1--9.

\bibitem{Ahmad2009a}
A.~Ahmad, B.~Krill, A.~Amira, and H.~Rabah, ``{3D Haar wavelet transform with
  dynamic partial reconfiguration for 3D medical image compression},'' in
  \emph{{Proc. IEEE Biomedical Circuits and Systems Conf. BioCAS 2009}}, 2009,
  pp. 137--140.

\bibitem{Dunham2009}
M.~E. Dunham, Z.~Baker, M.~Stettler, M.~Pigue, P.~Graham, E.~N. Schmierer, and
  J.~Power, ``{High Efficiency Space-Based Software Radio Architectures: A
  Minimum Size, Weight, and Power TeraOps Processor},'' in \emph{{Proc. Int.
  Conf. Reconfigurable Computing and FPGAs ReConFig '09}}, 2009, pp. 326--331.

\bibitem{SDCDPR6645549}
S.~{Di Carlo}, G.~{Gambardella}, M.~{Indaco}, P.~{Prinetto}, D.~{Rolfo}, and
  P.~{Trotta}, ``Dependable dynamic partial reconfiguration with minimal area
  time overheads on xilinx fpgas,'' in \emph{2013 23rd International Conference
  on Field programmable Logic and Applications}, Sep. 2013, pp. 1--4.

\bibitem{7169365}
M.~Doma{\'n}ski, J.~Konieczny, M.~Kurc, A.~{\L}uczak, J.~Siast, O.~Stankiewicz,
  and K.~Wegner, ``{Fast depth estimation on mobile platforms and FPGA
  devices},'' in \emph{{2015 3DTV-Conference: The True Vision - Capture,
  Transmission and Display of 3D Video (3DTV-CON)}}, July 2015, pp. 1--4.

\bibitem{7406067}
H.~Sun and Q.~Ding, ``{The Design of the Mobile Phone Module Information
  Encryption System Based on FPGA},'' in \emph{{2015 Fifth International
  Conference on Instrumentation and Measurement, Computer, Communication and
  Control (IMCCC)}}, Sep. 2015, pp. 1343--1347.

\bibitem{Perera:2015:AFR:2889287}
D.~G. Perera, ``Analysis of fpga-based reconfiguration methods for mobile and
  embedded applications,'' in \emph{Proceedings of the 12th FPGAworld
  Conference 2015}.\hskip 1em plus 0.5em minus 0.4em\relax ACM, 2015, pp.
  15--20.

\bibitem{aqtr2018}
A.~{Carelli}, C.~A. {Cristofanini}, A.~{Vallero}, C.~{Basile}, P.~{Prinetto},
  and S.~{Di Carlo}, ``Securing bitstream integrity, confidentiality and
  authenticity in reconfigurable mobile heterogeneous systems,'' in \emph{2018
  IEEE International Conference on Automation, Quality and Testing, Robotics
  (AQTR)}, May 2018, pp. 1--6.

\bibitem{BrickellCC04}
E.~F. Brickell, J.~Camenisch, and L.~Chen, ``{Direct anonymous attestation.}''
  in \emph{{ACM Conference on Computer and Communications Security}},
  V.~Atluri, B.~Pfitzmann, and P.~D. McDaniel, Eds.\hskip 1em plus 0.5em minus
  0.4em\relax ACM, 2004, pp. 132--145.

\bibitem{TPMSPECS}
\BIBentryALTinterwordspacing
{Trusted Computing Group}. {Trusted Platform Module 1.2 Main Specification}.
  [Online]. Available:
  \url{https://trustedcomputinggroup.org/resource/tpm-main-specification/}
\BIBentrySTDinterwordspacing

\bibitem{1027797}
C.~S. Collberg and C.~Thomborson, ``{Watermarking, tamper-proofing, and
  obfuscation - tools for software protection},'' \emph{IEEE Transactions on
  Software Engineering}, vol.~28, no.~8, pp. 735--746, Aug 2002.

\bibitem{1212692}
G.~Naumovich and N.~Memon, ``{Preventing piracy, reverse engineering, and
  tampering},'' \emph{Computer}, vol.~36, no.~7, pp. 64--71, July 2003.

\bibitem{SDCSW5749894}
P.~{Falcarin}, S.~{Di Carlo}, A.~{Cabutto}, N.~{Garazzino}, and D.~{Barberis},
  ``Exploiting code mobility for dynamic binary obfuscation,'' in \emph{2011
  World Congress on Internet Security (WorldCIS-2011)}, Feb 2011, pp. 114--120.

\bibitem{SDCSW5727971}
C.~{Basile}, S.~{Di Carlo}, and A.~{Scionti}, ``Fpga-based remote-code
  integrity verification of programs in distributed embedded systems,''
  \emph{IEEE Transactions on Systems, Man, and Cybernetics, Part C
  (Applications and Reviews)}, vol.~42, no.~2, pp. 187--200, March 2012.

\bibitem{Abraham:1991fk}
\BIBentryALTinterwordspacing
D.~G. Abraham, G.~M. Dolan, G.~P. Double, and J.~V. Stevens, ``{Transaction
  security system},'' \emph{IBM Syst. J.}, vol.~30, no.~2, pp. 206--229, Mar.
  1991. [Online]. Available: \url{http://dx.doi.org/10.1147/sj.302.0206}
\BIBentrySTDinterwordspacing

\bibitem{Wollinger2004}
T.~Wollinger, J.~Guajardo, and C.~Paar, ``{Security on FPGAs: State-of-the-art
  implementations and attacks},'' \emph{ACM Trans. Embed. Comput. Syst.},
  vol.~3, no.~3, pp. 534--574, 2004.

\bibitem{6881481}
S.~Trimberger and J.~Moore, ``{FPGA security: From features to capabilities to
  trusted systems},'' in \emph{{2014 51st ACM/EDAC/IEEE Design Automation
  Conference (DAC)}}, June 2014, pp. 1--4.

\bibitem{7238102}
R.~Druyer, L.~Torres, P.~Benoit, P.~V. Bonzom, and P.~Le-Quere, ``{A survey on
  security features in modern FPGAs},'' in \emph{{2015 10th International
  Symposium on Reconfigurable Communication-centric Systems-on-Chip
  (ReCoSoC)}}, June 2015, pp. 1--8.

\bibitem{Note:2008qf}
\BIBentryALTinterwordspacing
J.-B. Note and {\'E}.~Rannaud, ``{From the bitstream to the netlist},'' in
  \emph{{ACM/SIGDA Symposium on Field Programmable Gate Arrays}}.\hskip 1em
  plus 0.5em minus 0.4em\relax ACM New York, NY, USA, February 2008, pp.
  264--264. [Online]. Available: \url{http://islsm.org/~jb/debit/bitstream.pdf}
\BIBentrySTDinterwordspacing

\bibitem{Moradi:2011}
\BIBentryALTinterwordspacing
A.~Moradi, A.~Barenghi, T.~Kasper, and C.~Paar, ``{On the Vulnerability of FPGA
  Bitstream Encryption Against Power Analysis Attacks: Extracting Keys from
  Xilinx Virtex-II FPGAs},'' in \emph{{Proceedings of the 18th ACM Conference
  on Computer and Communications Security}}, ser. {CCS '11}.\hskip 1em plus
  0.5em minus 0.4em\relax New York, NY, USA: ACM, 2011, pp. 111--124. [Online].
  Available: \url{http://doi.acm.org/10.1145/2046707.2046722}
\BIBentrySTDinterwordspacing

\bibitem{Moradi:2013}
\BIBentryALTinterwordspacing
A.~Moradi, D.~Oswald, C.~Paar, and P.~Swierczynski, ``{Side-channel Attacks on
  the Bitstream Encryption Mechanism of Altera Stratix II: Facilitating
  Black-box Analysis Using Software Reverse-engineering},'' in
  \emph{{Proceedings of the ACM/SIGDA International Symposium on Field
  Programmable Gate Arrays}}, ser. {FPGA '13}.\hskip 1em plus 0.5em minus
  0.4em\relax New York, NY, USA: ACM, 2013, pp. 91--100. [Online]. Available:
  \url{http://doi.acm.org/10.1145/2435264.2435282}
\BIBentrySTDinterwordspacing

\bibitem{7857187}
R.~Karam, T.~Hoque, S.~Ray, M.~Tehranipoor, and S.~Bhunia, ``{Robust bitstream
  protection in FPGA-based systems through low-overhead obfuscation},'' in
  \emph{{2016 International Conference on ReConFigurable Computing and FPGAs
  (ReConFig)}}, Nov 2016, pp. 1--8.

\bibitem{drimer2007}
S.~Drimer, ``{Authenticated of FPGA bitstreams: why and how},'' \emph{In
  Applied Reconfigurable Computing}, vol. 4419, pp. 73--84, 2007.

\bibitem{6482382}
T.~Thanh, V.~H. Tiep, T.~H. Vu, P.~N. Nam, and N.~V. Cuong, ``{Secure remote
  updating of bitstream in partial reconfigurable embedded systems based on
  FPGA},'' in \emph{{2013 International Conference on Computing, Management and
  Telecommunications (ComManTel)}}, Jan 2013, pp. 152--156.

\bibitem{Stigge:2006nx}
M.~Stigge, H.~Platz, W.~Muller, and J.-P. Redlich, ``{Reversing CRC theory and
  practice},'' Humboldt University Berlin, Technical Report SAR-PR-2006-05,
  2006.

\bibitem{Parelkar:2005qe}
M.~M. Parelkar and K.~Gaj, ``{Implementation of EAX mode of operation for FPGA
  bitstream encryption and authentication},'' in \emph{{Proc. IEEE Int
  Field-Programmable Technology Conf}}, 2005, pp. 335--336.

\bibitem{Parelkar:2005kl}
M.~M. Parelkar, ``{Authenticated encryption in hardware},'' Master's thesis,
  George Mason University, 2005.

\bibitem{Badrignans:2010qf}
\BIBentryALTinterwordspacing
B.~Badrignans, D.~Champagne, R.~Elbaz, C.~Gebotys, and L.~Torres, ``{SARFUM:
  Security Architecture for Remote FPGA Update and Monitoring},'' \emph{ACM
  Trans. Reconfigurable Technol. Syst.}, vol.~3, no.~2, pp. 8:1--8:29, May
  2010. [Online]. Available: \url{http://doi.acm.org/10.1145/1754386.1754389}
\BIBentrySTDinterwordspacing

\bibitem{Bossuet2004}
L.~Bossuet, G.~Gogniat, and W.~Burleson, ``{Dynamically configurable security
  for SRAM FPGA bitstreams},'' in \emph{{Proc. 18th Int. Parallel and
  Distributed Processing Symp}}, 2004.

\bibitem{Hori2008}
Y.~Hori, A.~Satoh, H.~Sakane, and K.~Toda, ``{Bitstream encryption and
  authentication with AES-GCM in dynamically reconfigurable systems},'' in
  \emph{{Proc. Int. Conf. Field Programmable Logic and Applications FPL 2008}},
  2008, pp. 23--28.

\bibitem{Kashyap:2016}
\BIBentryALTinterwordspacing
H.~Kashyap and R.~Chaves, ``{Compact and On-the-Fly Secure Dynamic
  Reconfiguration for Volatile FPGAs},'' \emph{ACM Trans. Reconfigurable
  Technol. Syst.}, vol.~9, no.~2, pp. 11:1--11:22, Jan. 2016. [Online].
  Available: \url{http://doi.acm.org/10.1145/2816822}
\BIBentrySTDinterwordspacing

\bibitem{6029984}
R.~Maes, D.~Schellekens, and I.~Verbauwhede, ``{A Pay-per-Use Licensing Scheme
  for Hardware IP Cores in Recent SRAM-Based FPGAs},'' \emph{IEEE Transactions
  on Information Forensics and Security}, vol.~7, no.~1, pp. 98--108, Feb 2012.

\bibitem{7987601}
S.~K. K., S.~Sahoo, A.~Mahapatra, A.~K. Swain, and K.~K. Mahapatra, ``{A
  Flexible Pay-per-Device Licensing Scheme for FPGA IP Cores},'' in \emph{{2017
  IEEE Computer Society Annual Symposium on VLSI (ISVLSI)}}, July 2017, pp.
  677--682.

\bibitem{7168561}
L.~Zhang and C.~Chang, ``{Public key protocol for usage-based licensing of FPGA
  IP cores},'' in \emph{{2015 IEEE International Symposium on Circuits and
  Systems (ISCAS)}}, May 2015, pp. 25--28.

\bibitem{TPM}
\BIBentryALTinterwordspacing
{Trusted Computing Group}. Trusted platform module library part 3: Commands
  (level 00 revision 01.38). [Online]. Available:
  \url{https://www.trustedcomputinggroup.org/wp-content/uploads/TPM-Rev-2.0-Part-3-Commands-01.38.pdf}
\BIBentrySTDinterwordspacing

\bibitem{4439246}
T.~Guneysu, B.~Moller, and C.~Paar, ``{Dynamic Intellectual Property Protection
  for Reconfigurable Devices},'' in \emph{{2007 International Conference on
  Field-Programmable Technology}}, Dec 2007, pp. 169--176.

\bibitem{7031902}
J.~Zhang, Y.~Lin, Y.~Lyu, and G.~Qu, ``{A PUF-FSM Binding Scheme for FPGA IP
  Protection and Pay-Per-Device Licensing},'' \emph{IEEE Transactions on
  Information Forensics and Security}, vol.~10, no.~6, pp. 1137--1150, June
  2015.

\bibitem{7733105}
D.~Amelino, M.~Barbareschi, and A.~Cilardo, ``{An IP Core Remote Anonymous
  Activation Protocol},'' \emph{IEEE Transactions on Emerging Topics in
  Computing}, vol.~6, no.~2, pp. 258--268, April 2018.

\bibitem{chen2008pairing}
L.~Chen, P.~Morrissey, and N.~P. Smart, ``Pairings in trusted computing,'' in
  \emph{Pairing-Based Cryptography -- Pairing 2008}, S.~D. Galbraith and K.~G.
  Paterson, Eds.\hskip 1em plus 0.5em minus 0.4em\relax Berlin, Heidelberg:
  Springer Berlin Heidelberg, 2008, pp. 1--17.

\bibitem{Brickell2009SSN}
\BIBentryALTinterwordspacing
E.~Brickell, L.~Chen, and J.~Li, ``Simplified security notions of direct
  anonymous attestation and a concrete scheme from pairings,'' \emph{Int. J.
  Inf. Secur.}, vol.~8, no.~5, pp. 315--330, Sep. 2009. [Online]. Available:
  \url{http://dx.doi.org/10.1007/s10207-009-0076-3}
\BIBentrySTDinterwordspacing

\bibitem{Chen10efficient}
\BIBentryALTinterwordspacing
L.~Chen, ``A {DAA} scheme requiring less {TPM} resources,'' \emph{{IACR}
  Cryptology ePrint Archive}, vol. 2010, p.~8, 2010. [Online]. Available:
  \url{http://eprint.iacr.org/2010/008}
\BIBentrySTDinterwordspacing

\bibitem{Chen2010efficient2}
L.~Chen, D.~Page, and N.~P. Smart, ``On the design and implementation of an
  efficient daa scheme,'' in \emph{Smart Card Research and Advanced
  Application}, D.~Gollmann, J.-L. Lanet, and J.~Iguchi-Cartigny, Eds.\hskip
  1em plus 0.5em minus 0.4em\relax Berlin, Heidelberg: Springer Berlin
  Heidelberg, 2010, pp. 223--237.

\bibitem{Camenisch20116SDH}
\BIBentryALTinterwordspacing
J.~Camenisch, M.~Drijvers, and A.~Lehmann, ``Anonymous attestation using the
  strong diffie hellman assumption revisited,'' in \emph{Trust and Trustworthy
  Computing - 9th International Conference, {TRUST} 2016, Vienna, Austria,
  August 29-30, 2016, Proceedings}, 2016, pp. 1--20. [Online]. Available:
  \url{https://doi.org/10.1007/978-3-319-45572-3\_1}
\BIBentrySTDinterwordspacing

\bibitem{Camenisch2016UCD}
\BIBentryALTinterwordspacing
------, ``Universally composable direct anonymous attestation,'' in
  \emph{Proceedings, Part II, of the 19th IACR International Conference on
  Public-Key Cryptography --- PKC 2016 - Volume 9615}.\hskip 1em plus 0.5em
  minus 0.4em\relax New York, NY, USA: Springer-Verlag New York, Inc., 2016,
  pp. 234--264. [Online]. Available:
  \url{http://dx.doi.org/10.1007/978-3-662-49387-8_10}
\BIBentrySTDinterwordspacing

\bibitem{menezes}
A.~J. Menezes, S.~A. Vanstone, and P.~C.~V. Oorschot, \emph{Handbook of Applied
  Cryptography}, 1st~ed.\hskip 1em plus 0.5em minus 0.4em\relax Boca Raton, FL,
  USA: CRC Press, Inc., 1996.

\bibitem{secubeds}
\BIBentryALTinterwordspacing
{Blu5 View}. {SEcube(tm) Data Sheet Rev. 7 DUI 15082DS}. [Online]. Available:
  \url{https://www.secube.eu/site/assets/files/1145/secube_datasheet_-_r7.pdf}
\BIBentrySTDinterwordspacing

\bibitem{miracllib}
\BIBentryALTinterwordspacing
M.~Scott, ``Miracl-a multiprecision integer and rational arithmetic c/c++
  library,'' \emph{http://www.shamus.ie}, 2003. [Online]. Available:
  \url{https://ci.nii.ac.jp/naid/10026809141/en/}
\BIBentrySTDinterwordspacing

\bibitem{PCIDSS}
P.~Council, ``{PCI DSS Requirements and Security Assessment Procedures, version
  3.2.(2016)},'' Tech. Rep., 2016.

\bibitem{rfc5246}
T.~Dierks and {E. Rescorla}, ``{The Transport Layer Security (TLS) Protocol
  Version 1.2},'' Internet Requests for Comments, IETF, {RFC} 5246, 2008.

\bibitem{ipsec}
S.~Kent and R.~Atkinson, ``Security architecture for the internet protocol,''
  Internet Requests for Comments, IETF, United States, {RFC} 2401, 1998.

\bibitem{8110067}
M.~Bollo, A.~Carelli, S.~Di~Carlo, and P.~Prinetto, ``{Side-channel analysis of
  SEcube(tm); platform},'' in \emph{{2017 IEEE East-West Design Test Symposium
  (EWDTS)}}, Sept 2017, pp. 1--5.

\bibitem{blackhat2010}
C.~Tarnovsky. (2010) Hacking the smartcard chip.
  \url{https://www.blackhat.com/html/bh-dc-10/bh-dc-10-briefings.html}.

\bibitem{217652}
\BIBentryALTinterwordspacing
S.~Han, W.~Shin, J.-H. Park, and H.~Kim, ``A bad dream: Subverting trusted
  platform module while you are sleeping,'' in \emph{27th {USENIX} Security
  Symposium ({USENIX} Security 18)}.\hskip 1em plus 0.5em minus 0.4em\relax
  Baltimore, MD: {USENIX} Association, aug 2018, pp. 1229--1246. [Online].
  Available:
  \url{https://www.usenix.org/conference/usenixsecurity18/presentation/han}
\BIBentrySTDinterwordspacing

\bibitem{244048}
\BIBentryALTinterwordspacing
D.~Moghimi, B.~Sunar, T.~Eisenbarth, and N.~Heninger, ``Tpm-fail: {TPM} meets
  timing and lattice attacks,'' in \emph{29th {USENIX} Security Symposium
  ({USENIX} Security 20)}.\hskip 1em plus 0.5em minus 0.4em\relax Boston, MA:
  {USENIX} Association, aug 2020. [Online]. Available:
  \url{https://www.usenix.org/conference/usenixsecurity20/presentation/moghimi}
\BIBentrySTDinterwordspacing

\end{thebibliography}


%

\begin{IEEEbiography}[{\includegraphics[width=1in,height=1.25in,clip,keepaspectratio]{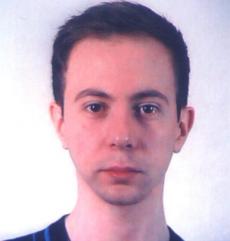}}]{Alberto Carelli}
(S'18) received a M.Sc. in computer engineering from Politecnico di Torino in Italy. 
Currently he is a Ph.D. student at the Department of Control and Computer Engineering of Politecnico di Torino in Italy. 
His research interests focus on hardware security and reliability for cyber-physical systems and critical infrastructures. 
\end{IEEEbiography}

\begin{IEEEbiography}[{\includegraphics[width=1in,height=1.25in,clip,keepaspectratio]{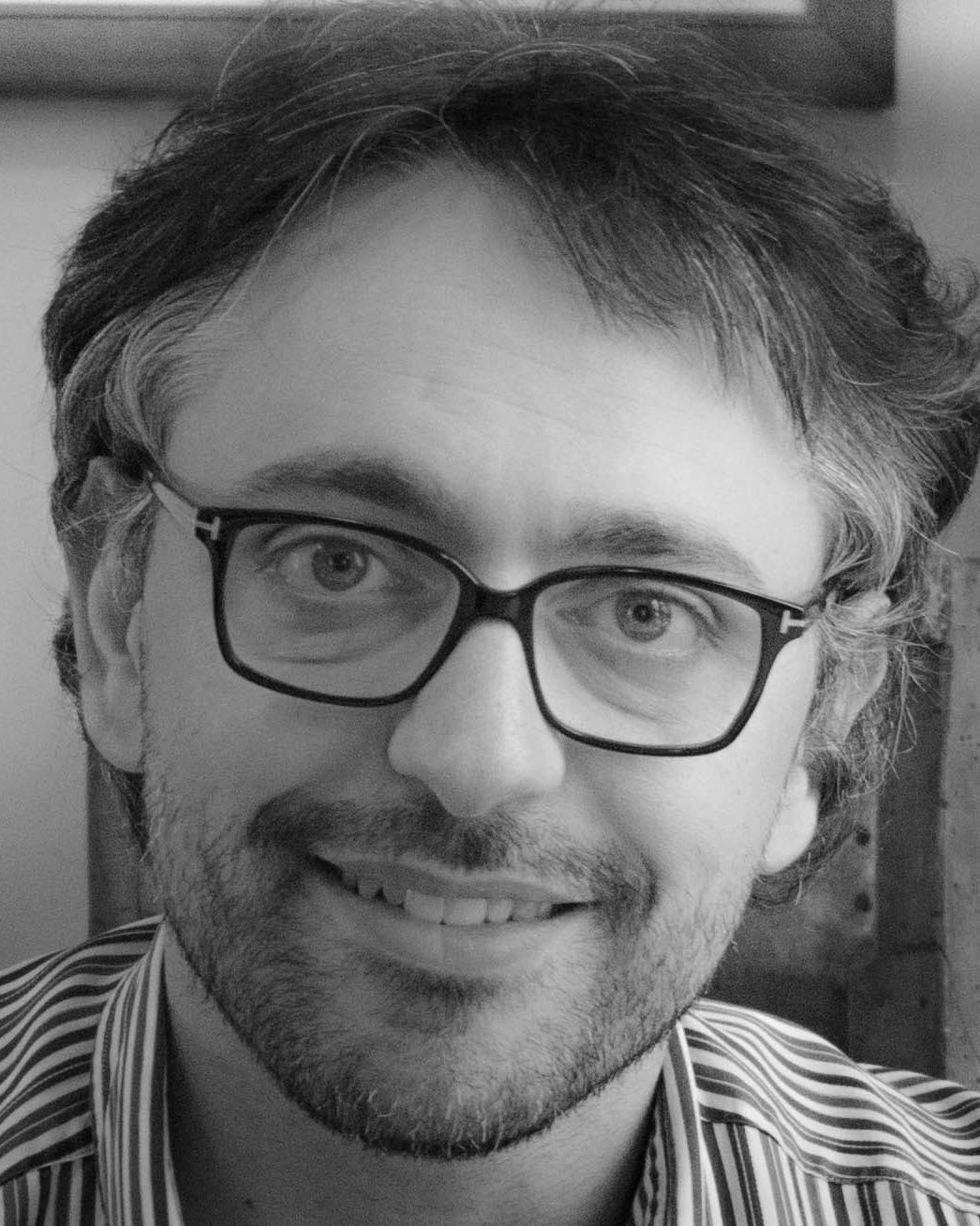}}]{Cataldo Basile} received a M.Sc. (summa cum laude) in 2001 and a Ph.D. in Computer Engineering in 2005 from the Politecnico di Torino, where is currently a assistant professor. His research is mainly concerned with software security, software attestation, policy-based management of security in physical and virtual networked environments, and general models for detection, resolution and reconciliation of specification conflicts.
\end{IEEEbiography}

\begin{IEEEbiography}[{\includegraphics[width=1in,height=1.25in,clip,keepaspectratio]{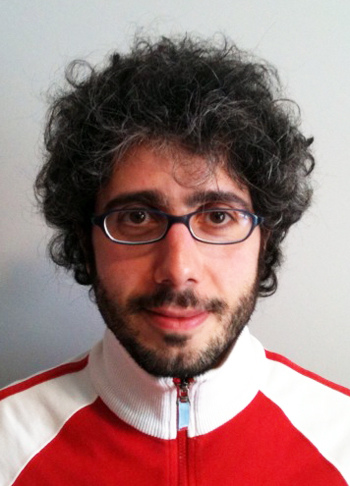}}]{Alessandro Savino}
(M'14) received a Ph.D. in information technologies and a M.Sc. degree in computer engineering from Politecnico di Torino, Italy, where he is an assistant professor in the Department of Control and Computer Engineering. His main research topics are microprocessor test and software-based self-test as well as bioinformatics and image processing.
\end{IEEEbiography}

\begin{IEEEbiography}[{\includegraphics[width=1in,height=1.25in,clip,keepaspectratio]{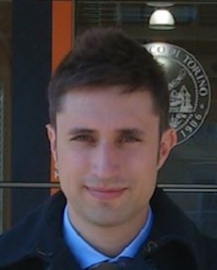}}]{Alessandro Vallero}
(S'15) received a Ph.D. in computer engineering from Politecnico di Torino in Italy and a M.Sc. degree in electronic engineering from the University of Illinois at Chicago, US, and Politecnico di Torino, Italy. Currently he is a postdoc at the Department of Control and Computer Engineering of Politecnico di Torino in Italy. His research interests focus on system level reliability and reliable reconfigurable systems. 
\end{IEEEbiography}

\begin{IEEEbiography}[{\includegraphics[width=1in,height=1.25in,clip,keepaspectratio]{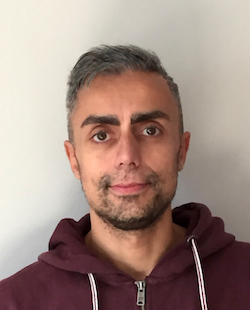}}]{Stefano Di Carlo}
(S'00-M'03-SM'11) received a M.Sc. degree in computer engineering and a Ph.D. degree in information technologies from Politecnico di Torino, Italy, where he is a tenured Associate professor. 
His research interests include DFT, BIST, and dependability. He has coordinated the EU-FP7 CLERECO on Cross-Layer Early Reliability Estimation for the Computing cOntinuum. 
Di Carlo has published more than 170 papers in peer reviewed IEEE and ACM journals and conferences. 
He regularly serves on the Organizing and Program Committees of major IEEE and ACM conferences. He is a golden core member of the IEEE Computer Society and a senior member of the IEEE.
\end{IEEEbiography}



\vfill
\vfill

\end{document}